# A system capable of verifiably and privately screening global DNA synthesis

Carsten Baum[1,2], Jens Berlips[3], Walther Chen[3], Helena Cozzarini[3], Hongrui Cui[4], Ivan Damgard[1,*], Jiangbin Dong[5], Kevin M. Esvelt[3,6,*], Leonard Foner[3], Mingyu Gao[5,12], Dana Gretton[3,6], Martin Kysel[3], Juanru Li[4], Xiang Li[5], Omer Paneth[7], Ronald L. Rivest[7], Francesca Sage-Ling[3], Adi Shamir[8], Yue Shen[10], Meicen Sun[11], Vinod Vaikuntanathan[7], Lynn Van Hauwe[3], Theia Vogel[3], Benjamin Weinstein-Raun[3], Yun Wang[10], Daniel Wichs[9], Stephen Wooster[3], Andrew C. Yao[3,5,12,*], Yu Yu[4,12], Haoling Zhang[10], and Kaiyi Zhang[4]

[1]Department of Computer Science, Aarhus University, Denmark
[2]DTU Compute, Technical University of Denmark, Denmark
[3]SecureDNA Foundation, Switzerland
[4]Department of Computer Science and Engineering, Shanghai Jiao Tong University, China
[5]Institute for Interdisciplinary Information Sciences, Tsinghua University, China
[6]Media Lab, Massachusetts Institute of Technology, USA
[7]Computer Science and Artificial Intelligence Laboratory, Massachusetts Institute of Technology, USA
[8]Department of Applied Mathematics, Weizmann Institute of Science, Israel
[9]Department of Computer Science, Northeastern University, USA
[10]China National GeneBank, China
[11]Department of Political Science, Massachusetts Institute of Technology, USA
[12]Shanghai Qi Zhi Institute, China
*ivan@cs.au.dk
*andrewcyao@mail.tsinghua.edu.cn
*esvelt@mit.edu

## Summary

A free, automated system verifiably blocks hazardous DNA synthesis orders while triggering negligible false alarms and protecting customer privacy.

## Abstract

Printing custom DNA sequences is essential to scientific and biomedical research, but the technology can be used to manufacture plagues as well as cures. Just as ink printers recognize and reject attempts to counterfeit money, DNA synthesizers and assemblers should deny unauthorized requests to make viral DNA that could be misused. There are three complications. First, we don't need to quickly update printers to deal with newly discovered currencies, whereas we regularly learn of new potential pandemic viruses and other biological threats. Second, convincing counterfeit bills can't be printed in small pieces and taped together, while preventing the distributed synthesis and subsequent re-assembly of controlled sequences will require tracking which DNA fragments have been ordered across all providers and benchtop devices while protecting legitimate customer privacy. Finally, counterfeiting can at worst undermine faith in currency, whereas unauthorized DNA synthesis could be used to deliberately cause pandemics. Here we describe SecureDNA, a free, privacy-preserving, and fully automated system capable of verifiably screening all DNA synthesis orders of 30+ nucleotides against an up-to-date database of controlled sequences, and its operational performance and specificity when applied to 67 million nucleotides of DNA synthesized by providers in the United States, Europe, and China.

---

## Introduction

Custom DNA synthesis is foundational to biomedical research, underpinning everything from cancer immunotherapies to SARS-CoV-2 vaccines. However, the same technology can also be used to produce pathogens[1–3] (Fig. 1a). The first infectious virus to be assembled from synthetic DNA was generated in 2002 at a cost of more than $10 per nucleotide[4]. Just over two decades later, the price has fallen more than a thousandfold, the number of individuals with the necessary skills has grown from dozens to many thousands[5], and a pandemic has directly and indirectly killed over 20 million people[6], demonstrating the scale of harm that could be inflicted by a virus generated from synthetic DNA.

While no credible and accessible pandemic viruses are publicly known, numerous well-intentioned research programs aim to identify viruses capable of causing new pandemics and share their genome sequences[7–11]; one maintains a public list of viruses ranked by threat level[12]. Future discoveries and advances in biological programming will identify other pandemic-class agents that can be generated from synthetic DNA.

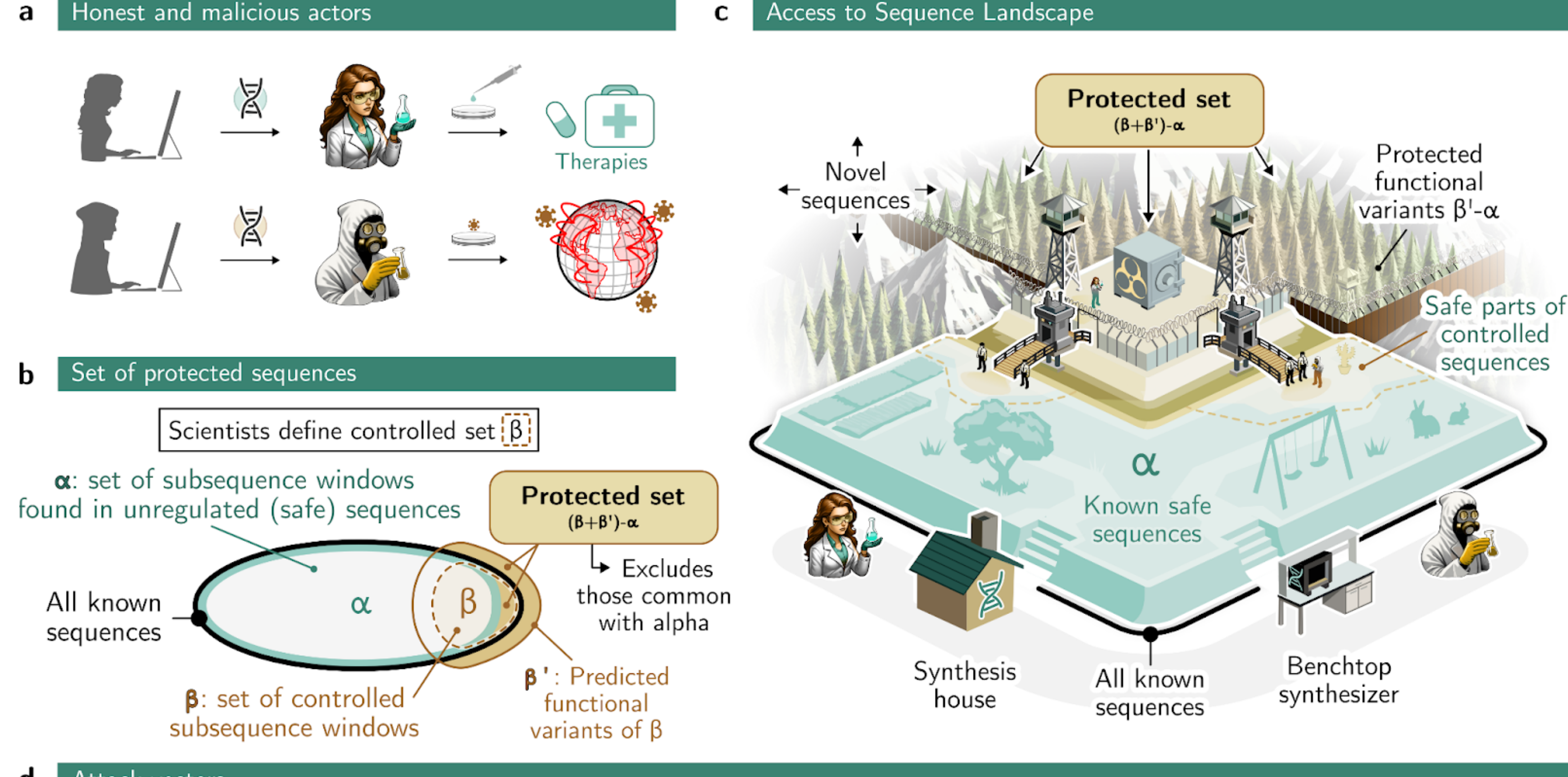

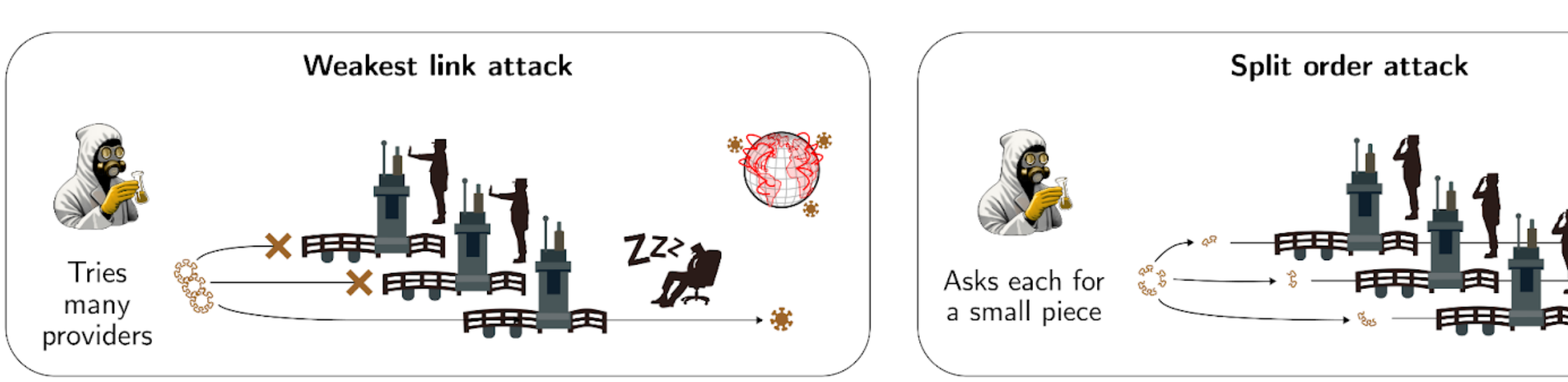

**Figure 1 | DNA synthesis screening ecosystem and security framework. a)** DNA synthesis enables therapeutic development and medical progress (top), but also presents biosafety and biosecurity risks (bottom). **b)** DNA screening framed as a set-membership problem using configurable sets. α represents constant-length subsequences found in non-controlled (safe) sequences, including most organisms and common molecular biology tools. Scientists and regulators define a safe set α, a controlled set β, and compute functional variants β', with the protected set (β+β')−α comprising subsequences from controlled sequences, or their variants, that do not appear in safe sequences. **c)** Screening must cover both paths to sequence space: synthesis houses and benchtop synthesizers. Orders containing protected-set sequences should require authorization, while sequences in α and novel sequences not in β' remain freely accessible. Ideal screening should only learn of any fragments in (β+β'). **d)** A weakest link attack bypasses screening by testing multiple providers to find one with inadequate protocols. A split order attack evades screening by splitting controlled sequences into fragments that are either too short to screen or appear harmless in isolation. Such attacks underscore the need for a globally adopted screening system that can detect all assembly-viable sequence lengths.

The obvious solution to the resulting proliferation problem, screening all synthetic DNA orders for a continuously updated list of controlled sequences, was first recommended in 2006[13]. Remarkably, the members of the International Gene Synthesis Consortium (IGSC) voluntarily monitor an estimated 80% of global DNA synthesis, even though the relative cost of human-assisted screening is rising with volume[14,15].

Nevertheless, synthesis screening remains an unsolved global problem:

- 36/38 providers recently shipped DNA fragments collectively sufficient to generate the 1918 pandemic influenza virus because they had no way to know that others were supplying the remaining fragments[16]
- Chatbots now suggest ways to evade screening, including ordering from non-IGSC providers[17]
- Oligonucleotides under 50 nucleotides can be assembled and used to generate pandemic threats[14]
- Scientific consensus on which sequences warrant control varies across regions and evolves over time
- Most benchtop devices do not screen, on-device screening is insecure, and cloud-based[18] is not private

The DNA industry is also at risk:

- Firms may be liable because they cannot verify that they performed best-in-practice screening
- Rising demand may overwhelm expert screening with false alarms and costs may become prohibitive[14]

Solving these problems requires a free, centralized, and privacy-preserving screening system configurable for any defined set of controlled sequences and effective for synthesis houses and benchtops (Fig. 1b-c).

Current screening systems rely on fuzzy-match algorithms that compare orders to both controlled and unregulated sequences, attempting to determine which they more closely resemble. This approach faces fundamental limitations: it incorrectly flags more and more unregulated sequences as the sequence window length decreases, it requires human expert review to resolve these ambiguous matches and consequently can't be used by benchtop DNA synthesis devices, it can be evaded by introducing mutations that make controlled sequences appear more similar to unregulated genes, and it cannot reliably detect split orders distributed across providers in different nations (Fig. 1d).

Overcoming these limitations requires a new approach that can address two design challenges:

1. *Bio-design*: translate the biological problem of sequence recognition into a computer science problem
2. *System-design*: implement an automated system capable of verifiably screening global DNA synthesis without delaying research or compromising customer or database privacy

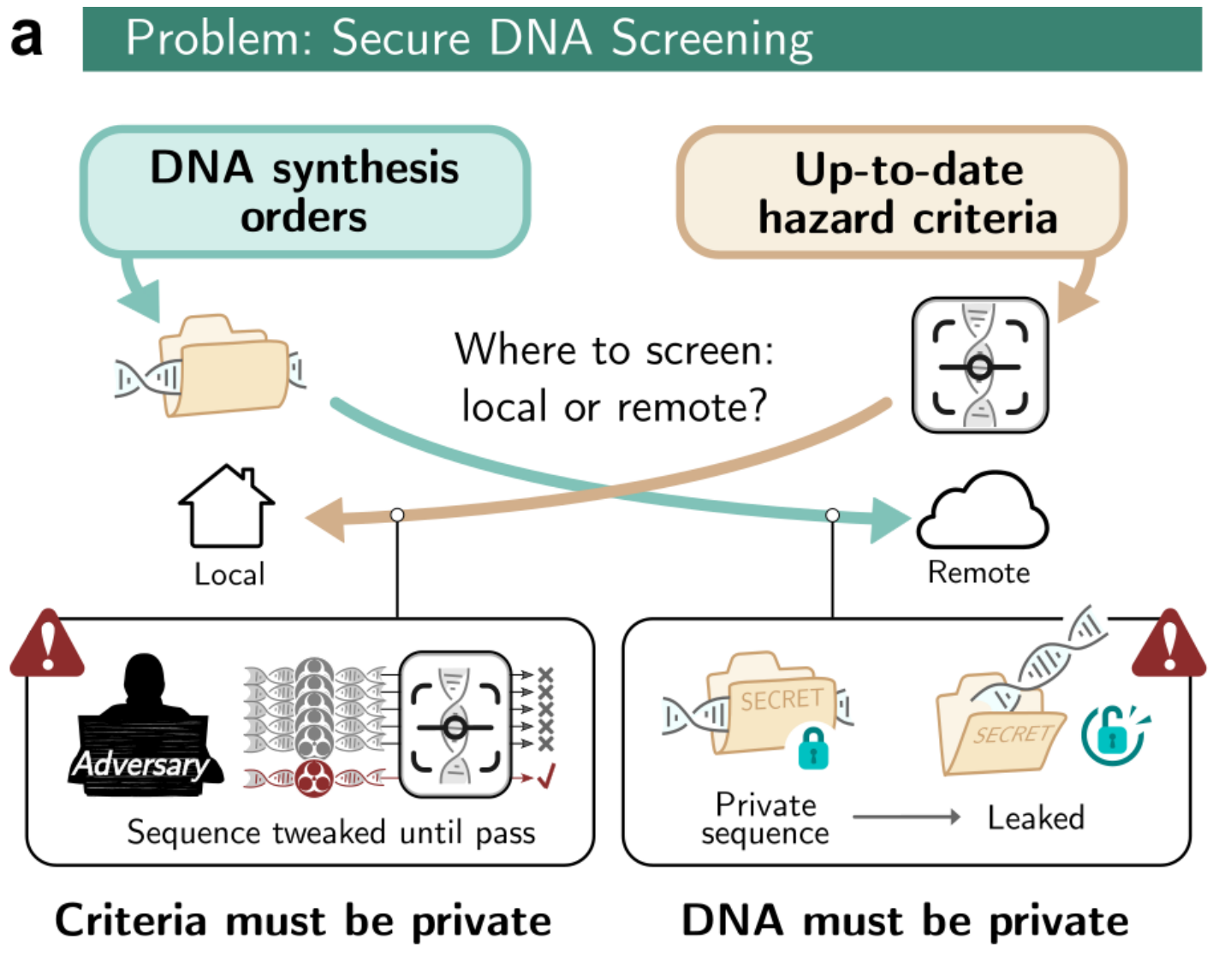

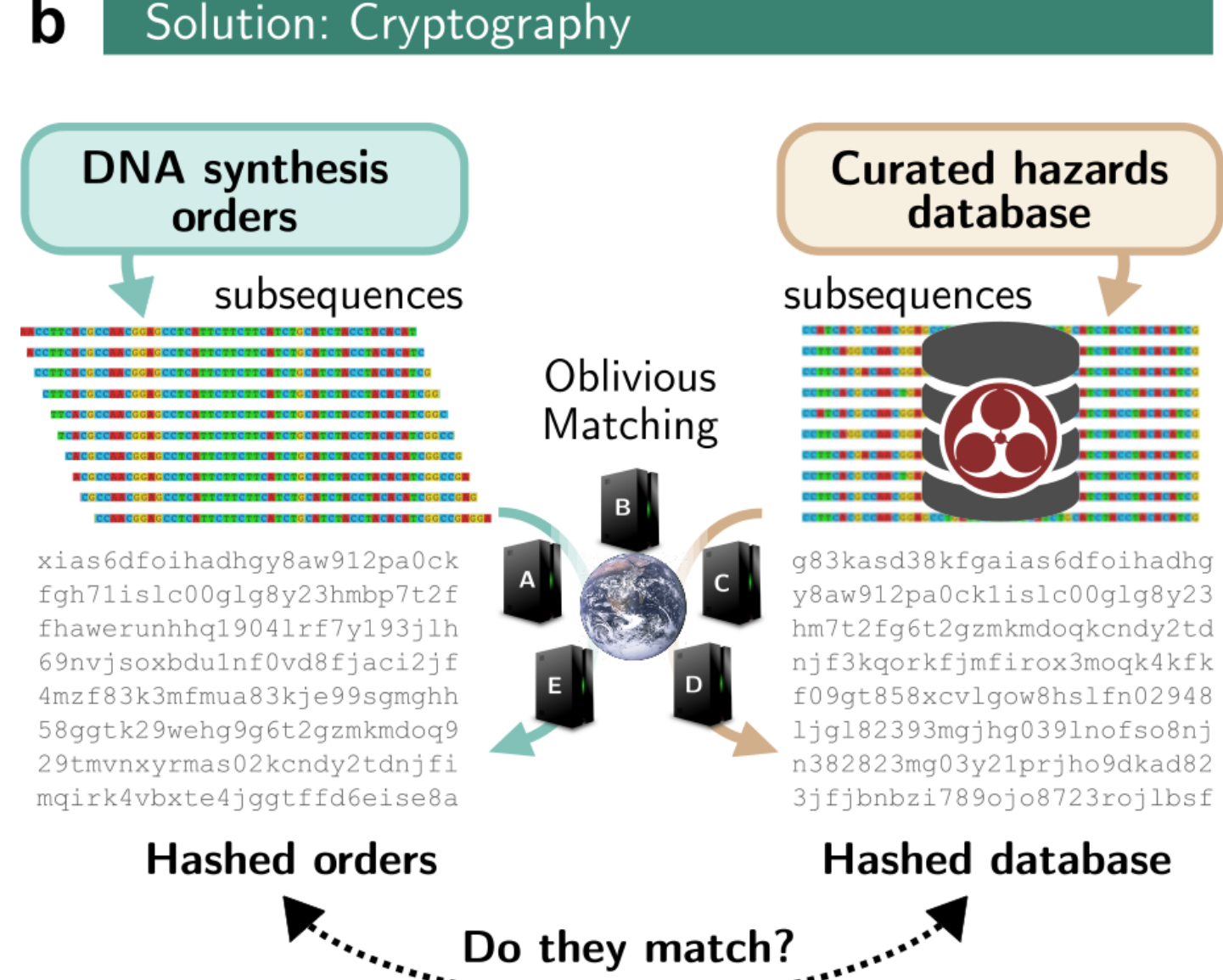

**Figure 2 | Securing DNA synthesis screening.**
**a)** Secure and universal DNA synthesis screening requires a way to verifiably determine whether DNA synthesis orders correspond to controlled biological functions, including controlled sequences split across multiple providers and benchtop devices, without disclosing anything else about the orders. Disclosing a private customer order may compromise trade secrets, while leaking the criteria makes it possible to evade screening by introducing mutations that make a controlled sequence appear to be unregulated. These constraints cannot be satisfied simultaneously.
**b)** The SecureDNA system allows synthesizers and database contributors to *obliviously* perform one-way transformations of their subsequence windows, which can be directly compared. The database provides the synthesizer with a timestamped verification that $n$ windows sent by the synthesizer were screened against a particular database version and can detect split order attacks if adopted by most providers.

In a companion paper[19], we describe a candidate solution to the first challenge: searching for exact matches to short subsequences that are unique to controlled genes and functional variants. By identifying these diagnostic fragments in advance and verifying that they don't appear in unregulated sequences, it can achieve perfect specificity among known sequences (Fig. 1b-c). This precision eliminates the need for human expert review, while the inclusion of functional variants renders screening robust against evasion attempts.

Here we address the system design challenge (Fig. 2) using a novel application of cryptography to enable centralized and privacy-preserving detection (Fig. 3). Our automated implementation, including a graphical user interface, screens at high speed and low cost while demonstrating very high specificity (Fig. 4), and implements a certificate system permitting authorized laboratories to seamlessly access controlled DNA (Fig. 5).

**Results**

*Theoretical crypto-design and analysis*

Suppose that a Customer places an order for sequence $s$ with a Synthesizer, who wants to know whether it's safe to make, without disclosing $s$—which may be a trade secret—to any eavesdroppers (Fig. 3a). The Synthesizer can ask an up-to-date remote Database $D$ of controlled DNA and peptide subsequences (Fig. 3b; SI Appendix A) to check whether any of the constant-length DNA and translated peptide subsequences found in $s$ are present in $D$. The crypto-design challenge is to find a way for the Database to answer without 1) learning anything consequential about $s$ or 2) revealing anything about $D$ beyond conveying the yes/no answer, even if some of the parties are compromised.

The simplest way for the Synthesizer and Database to privately determine whether any of their DNA or peptide subsequences are identical is to transform them using a one-way hash mapping $M$ and compare the hash outputs $M(s)$ and $M(D)$ (Fig. 3c). But either of them could rapidly enumerate all possible subsequences, letting them decrypt $s$ and $D$ (Fig. 3d), even when using a shared key $\kappa$ (Fig. 3e). To restore privacy, we introduce the Keyserver, who helps compute hashes using a frequently-changed key $\kappa$ while rate-limiting requests to match plausible DNA synthesis speeds, which renders enumeration attacks infeasible (Fig. 3f). The Synthesizer and Keyserver can compute hash outputs *obliviously* using cryptographic techniques: the Keyserver learns nothing about the input sequences or resulting hashes, while the Synthesizer learns only the hash outputs.

To screen for matches, the Database is given a table $H$ comprising the oblivious keyed hashes of all elements in the plaintext database $D$, without learning its contents. Whenever the Synthesizer submits a hashed query, the Database can convey whether there are any matches to $H$ without learning anything about the sequences apart from whether they match. Keeping a log of which subsequences have recently matched enables split-order detection. Note this approach is distinct from private set intersection, which is simultaneously unsuitable for some tasks demanded of screening (such as split order detection) and also far from being sufficiently performant (SI Appendix B).

The Database provides the Synthesizer with a timestamped signature verifying that screening was performed against a specific version of $H$, allowing providers to demonstrate compliance with legal requirements and minimize liability. Overall, this architecture ensures that no single party needs to know the list of controlled sequences or which subsequence windows are defended with functional variants—a critical feature, as knowledge of defended windows would enable evasion. In the future, the system could allow nations to screen for emerging threats without publicly acknowledging them[19].

To summarize the baseline system:

- Scientists and Keyserver build and update Database $H$ of obliviously hashed subsequences
- Customer orders sequence $s$ from Synthesizer
- Synthesizer and Keyserver obliviously hash subsequences
- Synthesizer sends hash outputs to Database
- Database tells Synthesizer if any are found in $H$
- If there are matches and Customer lacks authorization, Synthesizer refuses to make $s$

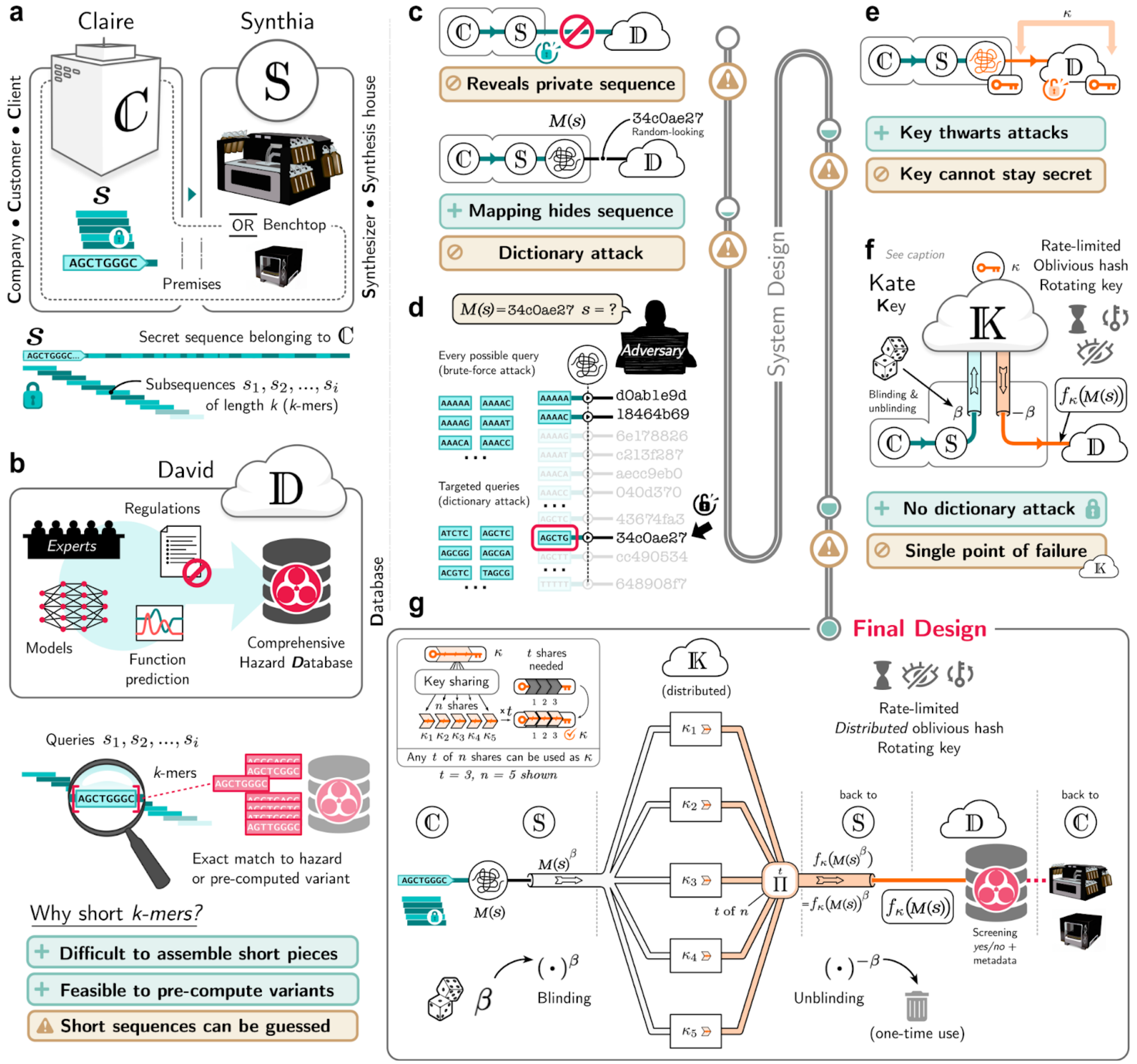

**Figure 3 | Cryptographic challenges and solutions for secure DNA synthesis screening**
**a)** Client Claire orders DNA from commercial synthesis house or benchtop Synthesizer Synthia. Claire's private DNA sequence $s$ is split into all subsequences of length $k$ ($k$-mers). **b)** In the simplest version, a centralized Database David is populated with $k$-mers from controlled sequences and predicted functional variants. Synthia sends $k$-mers to be screened, and David detects exact $k$-mer matches. If $k$ is small, David can screen small oligonucleotides that might otherwise be assembled into controlled genes. **c)** Synthia cannot send $s$ directly to David (as in current remote DNA screening systems) without violating Claire's privacy, but she could obscure $s$ with a cryptographic mapping $M(s)$ and compare to David's similarly hashed sequences. **d)** A standard hash is easily cracked when the input is short or has low entropy using a brute-force or dictionary attack (trying many possible inputs). **e)** A keyed hash with a secret key $\kappa$ can thwart dictionary attacks, but requires that $\kappa$ be known to both Synthia and David, in which case Synthia could interrogate $D$ and David could crack $s$. **f)** Keyserver Kate helps Synthia compute keyed hash $f_\kappa(M(s))$ with oblivious cryptographic hashing. Synthia "blinds" the sequence $s$ with a random value β before Kate applies $\kappa$, then unblinds it afterward, so no eavesdropper (including Kate) can learn $s$. Kate can limit the rate of evaluation of $f_\kappa(s)$, making dictionary attacks impossible. However, Kate is a single point of failure: an adversary compromising Kate gains $\kappa$, and if Kate is offline, global synthesis stops. **g)** In the final design, $f_\kappa(M(s))$ is distributed, making Kate's role robust. $\kappa$ is split into $n$ key shares among $n$ separated servers.

Evaluating $f_\kappa(M(s))$ requires a threshold number $t$ of the $n$ shares (n = 5, t = 3 shown). Up to ($n$-$t$) keyservers may be offline. $\kappa$ is rotated, e.g. biweekly. Any adversary must compromise $t$ servers simultaneously between key rotations to steal $\kappa$.

*Security analysis of baseline screening*

In the *semi-honest* model of cryptographic security, all parties follow the protocol, but may eavesdrop. Oblivious hashing ensures the Keyserver learns nothing about sequences or hashes, while the Synthesizer learns only whether matches exist—the minimum necessary for screening. If no controlled sequences are found, the Database learns only the total sequence length (which is biologically irrelevant) and possibly whether any subsequences are shared with other clients' queries. If $s$ contains controlled sequences, the Database learns which subsequences match ($\beta+\beta'$) from $D$, but cannot determine any other sequences in $s$ or $D$.

In the *malicious* model, a party may deviate from the protocol to eavesdrop or sabotage its execution. However, a fundamental premise is that the Synthesizer wants to avoid creating controlled sequences—no software can prevent a synthesizer that is not hardware-locked from generating controlled DNA. While a dishonest Synthesizer could disclose $s$, rate-limiting prevents them from reconstructing $D$ through repeated queries. A corrupt Database can disclose $H$ and any statistical correlations, but cannot interrogate $s$ or $D$ without help from the Keyserver—see SI Appendix B for a detailed analysis of information leakage. Keyserver could use $\kappa$ to hash all possible subsequences and thereby determine $s$ and $D$.

*Distributing the key*

The integrity of the Keyserver is so vital that we divide the key $\kappa$ into $n$ shares and distribute them across $n$ keyservers using Shamir secret sharing[20] (Fig. 3g). This architecture requires a threshold $t$ keyservers to jointly perform oblivious hashing; any smaller group learns nothing about $\kappa$. To prevent a malicious party from stealing $\kappa$ by gradually corrupting a majority of the keyservers, the system periodically updates key shares through a coordinated process with the same security guarantees as distributed oblivious hashing—no single party ever reconstructs $\kappa$. This achieves what cryptographers call *proactive security*: $\kappa$, and thereby all $s$ and $D$, remain protected even if all keyservers are eventually hacked as long as they are not compromised simultaneously. See SI Appendix C for the complete protocol.

*System design and implementation*

The database is constructed by partitioning controlled sequences into 30 or 42 nucleotide DNA windows and 20 amino acid peptide windows, which are tagged with geographical regions in which they are regulated. For a subset of windows, chosen pseudo-randomly, we stochastically include mutants and functional variants predicted using protein structure and sequence homology.

To ensure specificity among all known sequences, we check each candidate subsequence against a dataset of nucleotides and proteins derived from NCBI's nr/nt GenBank and GenPept databases and remove any that are also found in an unregulated sequence. This filtering process combined taxonomic analysis, keyword identification (e.g. "synthetic" or "recombinant"), and controlled sequence coverage metrics to improve database annotations. An analysis of controlled viral sequences revealed that nearly all contained thousands of k-mers ($k \geq 30$) unique to strains of the controlled virus, providing ample targets for effective screening.

The system can accommodate any arbitrarily defined set of controlled sequences, permitting customization by provider preference. By default, it detects controlled sequences defined in Australia Group, ITAR, or national regulations, viruses described as potential pandemic pathogens in the scientific literature, and viruses capable of human-to-human transmission or illness that could be misused. While the system returns an alert upon any unique match, it recommends denying synthesis for controlled viruses, potential pandemic viruses, genes encoding controlled toxins, and sequences that could confer controlled-pathogen capabilities to more accessible strains per the Australia Group's "endow or enhance pathogenicity" definition.

*Split order detection*

Malicious actors can evade screening by splitting controlled sequences into multiple individually harmless fragments and ordering them from different providers, who have no way of knowing whether others are making the remaining pieces[16]. Even if governments required logs of all controlled fragments ordered from their providers, nations would need to share data with one another to detect such attacks, which appears unlikely given diplomatic constraints. Because it is privacy-preserving, offered free of charge by a nonprofit in a neutral country, co-developed internationally, and can be used by centralized synthesis houses and benchtop synthesizers, SecureDNA could potentially be adopted widely enough to detect such attempts by monitoring which pieces have been ordered from different suppliers and notifying either suppliers or law enforcement of detected attacks, as appropriate. To enable split-order detection, the database stores hashes of controlled subsequences in order and logs which ones have been detected. However, the system learns nothing of which customer ordered which k-mers, only which provider they ordered it from.

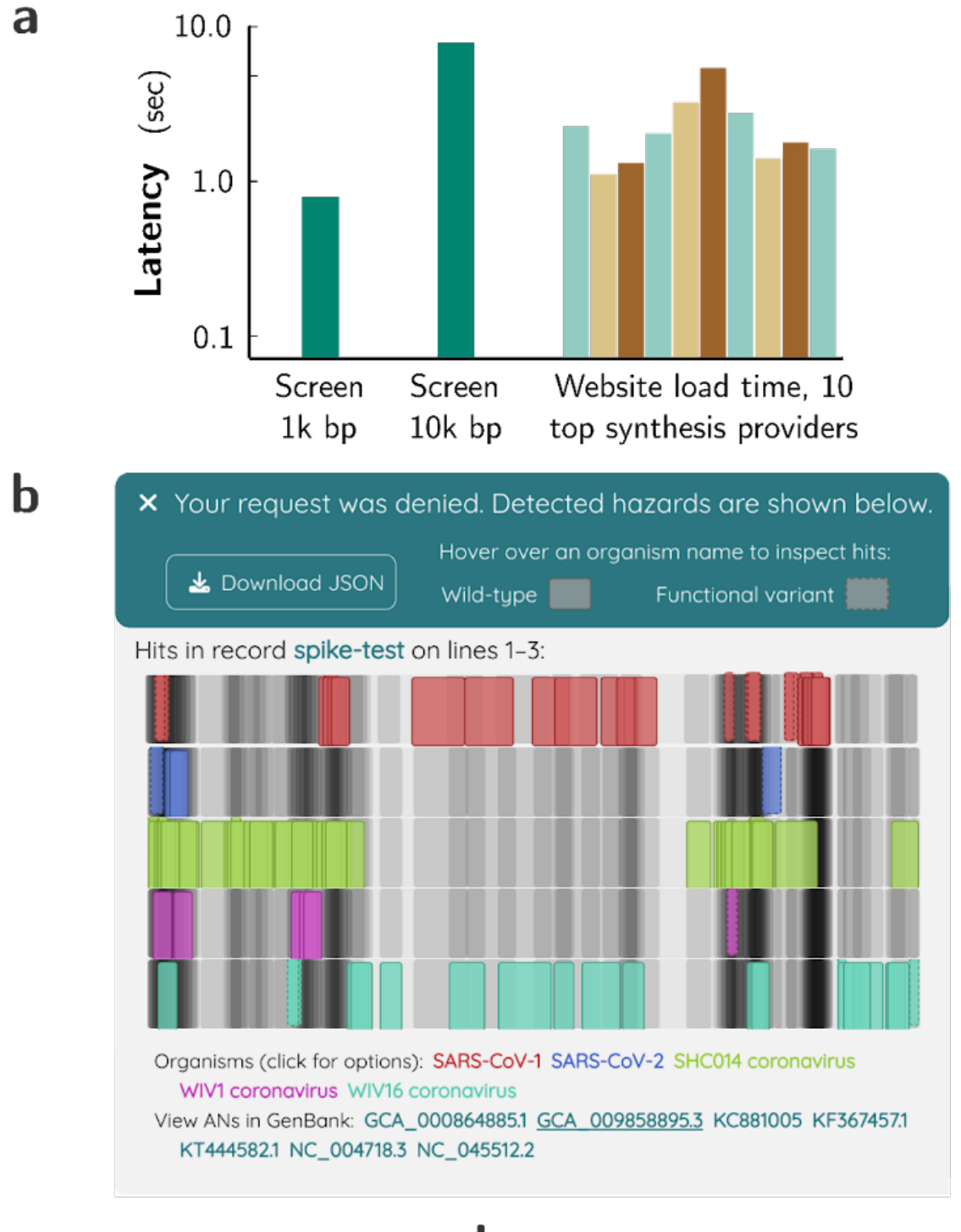

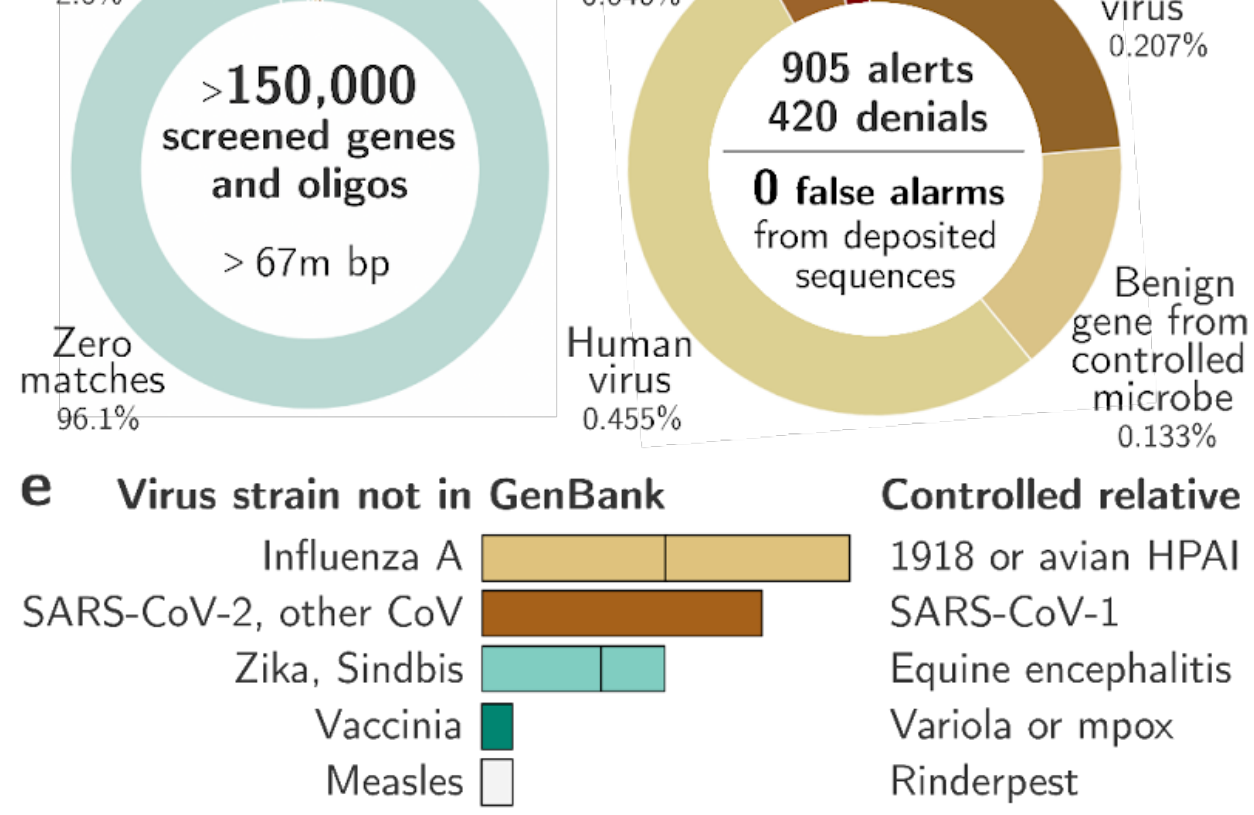

**Figure 4 | SecureDNA latency, output and specificity on real-world synthetic genes.** **a)** A comparison of the time required to screen 1,000- and 10,000-base-pair (bp) orders to the typical response times of randomly chosen DNA synthesis provider websites. SecureDNA does not meaningfully delay order placement and can return results to customers in real-time, unlike approaches that require expert human curation. **b)** SecureDNA returns matches to wild-type sequences and predicted functional variants of windows from public controlled pathogens (the intersection ofs and *D)* and potential pandemic viruses, and also flags sequences from human viruses and benign genes from regulated bacteria. A hover-over interface can view the windows matching each sequence separately or go to the relevant GenBank file. The depicted chimera was denied because it includes k-mers unique to SARS-CoV-1. **c)** Analyzing >67 million nucleotides of synthesized genes from multiple providers with SecureDNA flagged 0.57% of genes (yellow) and denied 0.27% (brown). Fragments present in both controlled (β or β') and safe (α) sequences are Curated, while sequences unique to β but often used in biotechnology (e.g. P2A peptides) are Common. Both represent matches but not alerts. **d)** Iterative manual BLAST revealed zero false alarms from known sequences: all 30 false alarms were from strains of related viruses not present in GenBank. **e)** Probable identities of the 30 false alarms (1 in 5,000 orders) flagged due to sequence windows not present in GenBank.

### *System operation*

Each synthesizer or provider running the open-source SecureDNA *synthclient* program converts FASTA files from sequence orders into all eligible subsequences, translations, and permutations. Screening proceeds in both forward and reverse-compliment directions across the query sequence[19] and, for benchtop synthesizers, includes base permutations to prevent evasion through reagent bottle swapping (Extended Data Fig. 1). This step negligibly reduces specificity and actually improves performance, as all 48 possible hash combinations (including all possible base permutations in both directions) are consolidated. After local processing, *synthclient* contacts the keyservers to perform oblivious hashing and sends the results to the database for comparison. Screening completes faster than most web pages load (Fig. 4a), providing an immediate 'accepted', 'alert', or 'denied' decision along with a signed, timestamped verification that screening was performed using a specific database version. The system also generates a visual mapping of any matches to public controlled sequences and notes relevant export control restrictions (Fig. 4b).

*Deployment and performance*

The production system operates with five keyservers distributed across multiple geographic zones to maximize fault tolerance, requiring any three to process requests. Under normal load, the system completes key rotation every 1-2 weeks, with the capacity to recompute the entire database $H$ in three days if needed. The primary computational bottleneck remains elliptic curve group operations, requiring 10-1000 microseconds per window depending on architecture.

Despite modest computational resource requirements (~$5-8K/year, Methods), the system demonstrates robust performance. *Synthclient* processes 1100 nucleotides per second per CPU thread, or about 2200 nt/second per hyperthreaded core, on provider hardware. Keyservers achieve 4400 nt/second on 4-thread CPUs. The database, using cloud SSD storage, handles 41,500 nt/second per disk, and over 74,000 nt/second in a physical machine with NVMe storage, with constant-time lookup regardless of database size up to 15x the size of the current database. Most providers need only a single $2,000 desktop computer or can screen for pennies per day using on-demand cloud servers, while benchtop synthesizers can operate with a $50 Raspberry Pi 4B (Extended Data Table 1). For customers with very large oligo-manufacturing needs, we have demonstrated 40 million nucleotides in under 10 minutes at a customer cost of 61 US cents for cloud-based clients. The current hardware configuration can screen more than 4 trillion nucleotides annually—sufficient for projected global synthesis volume through 2029.

As SecureDNA expands, we plan to increase security and redundancy by adding servers (SI Appendix D). Operating fees estimated at $15-30K/yr (depending on degree of redundancy) for servers housed in a colocation facility constitute the only recurring usage cost borne by SecureDNA (SI Appendix D), and are sufficient to support an order of magnitude more screening than the estimated size of the current worldwide market for both genes and oligos.

*Quantifying specificity using real-world synthesis data*

To measure specificity in a real-world context, we screened anonymized orders for over 150,000 genes and oligonucleotides synthesized by providers in the United States, Europe, and China, which together comprise over 61 million nucleotides generated for biological research and 6 million for DNA storage.

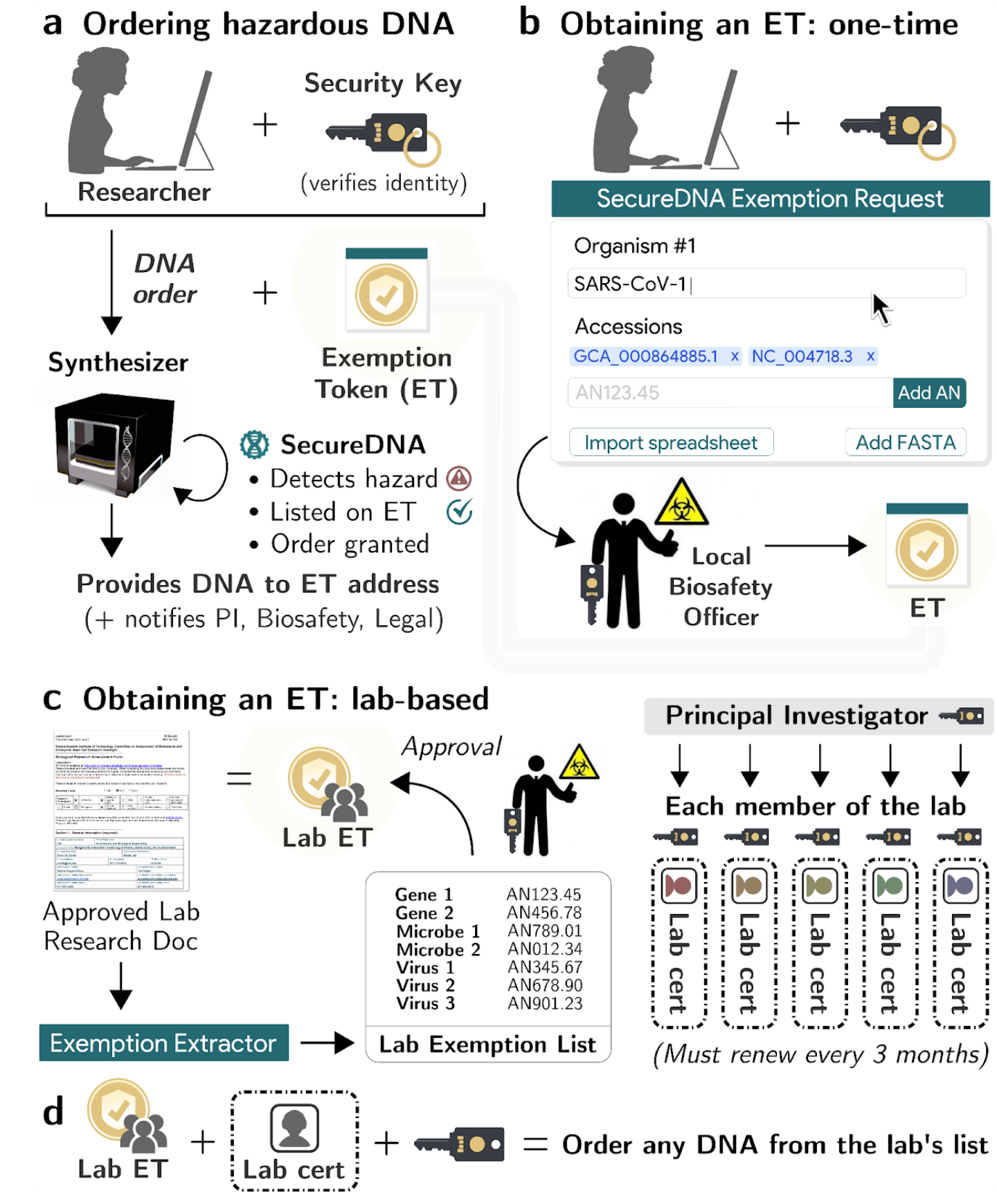

**Figure 5 | SecureDNA exemption lists for automated customer screening.**
**a)** Researchers with a hardware security key can submit an *exemption token* (ET) with their order to bypass denials for any controlled sequence listed on the ET**b)** To obtain a one-time ET, researchers can specify a controlled agent from a drop-down table, enter a GenBank accession number, or import a DNA sequence, then send it to their local biosafety officer to be approved and signed using the institution's certificate.**c)** Labs can obtain an ET that will work for all members by using an extraction tool on their research registration document to obtain a list of genes, microbes, and viruses listed in the document, which can then be approved by the biosafety officer. The lab principal investigator can grant 'lab certificates' to each member.**d)** A lab ET, lab cert, and matching security key can obtain DNA from any gene or organism the lab is approved to work with, including not-yet-deposited strains and predicted functional variants.

Over 99% of sequences passed screening, and an overwhelming majority (96%) featured zero matches. Among sequences with one or more matches (4%), two-thirds (2.6%) were found in unregulated sequences and were not flagged (Fig. 4c). One-third of those remaining (0.45%) were flagged as "Common" because they matched one of the many frequently used sequences in biotechnology that originated in a regulated organism or a human-infecting virus. Examples include the porcine teschovirus 2A peptide and the cytomegalovirus enhancer/promoter.

SecureDNA identified genes with at least one nucleotide or peptide match to a controlled organism or human virus in 0.85% of sequences (Fig. 4c). Each was investigated by running BLAST on the entire gene sequence, then on any subsequences that were not clearly identified in the first attempt, and manually labeling (Fig. 4d). Approximately half (0.455%) were from unregulated viruses capable of infecting humans, for which several providers have requested alerts; others (0.133%) matched benign genetic sequences from regulated microbes that could not confer pathogenicity upon more easily obtained related or avirulent strains.

Absent an exemption token (see below), we would have recommended denying just 0.272% of all ordered genes: 0.046% encoded a toxin or conferred pathogenicity upon a more easily obtained avirulent microbe, 0.207% were unique to a controlled or potential pandemic virus, and 0.019% (30 gene fragments) were from strains of unregulated human-infecting viruses not deposited in Genbank (Fig. 4d), all related to controlled viruses.

Crucially, we observed zero false alarms from known sequences in GenBank: the system perfectly distinguished between all known sequences. The 30 false alarms from sequences absent from GenBank were all from strains of related human-infecting viruses or deliberately created mutants that had never been deposited. We estimate that at current DNA synthesis rates, this false alarm rate would correspond to just ~4 wrongly denied sequences globally each day. By comparison, we anticipate approximately 50 legitimate authorization requests daily for controlled sequences, meaning more than 90% of all authorization requests would be for actual controlled sequences rather than false alarms. Nevertheless, it is imperative for authorized laboratories that work with these pathogens—or with controlled agents—to gain access to the DNA that they need to advance human knowledge and develop therapies, without delay.

*Automating customer screening and permissions*

Members of the International Gene Synthesis Consortium spend considerable time and effort screening their customers for legitimacy. In some cases, providers contact the customer's local biosafety authority to ask whether the customer is authorized to work with a particular controlled organism.

An ideal screening system would automatically grant researchers access to DNA corresponding to pathogens that their biosafety authority has already approved them to work with. SecureDNA allows researchers to submit an "exemption token" (ET) with their order to obtain any permitted controlled DNA (Fig. 5a). A one-time token contains a public key identification (PKI) certificate instructing the database to ignore matches corresponding to exempted controlled sequences, but only for requests originating from the user. This allows researchers to swiftly obtain DNA corresponding to any controlled gene, organism, or set of raw sequences by requesting an exemption from their biosafety officer, who approves it using their own security key and a certificate issued by a higher biosafety authority, such as one at the national level (Fig. 5b, Extended Data Fig. 2).

To facilitate seamless ordering, we wrote software to help biosafety authorities convert approved laboratory research registration documents into exemption tokens for permitted genes and organisms, then issue the lab an ET (Fig. 5c). The principal investigator can separately grant certificates to lab members, enabling them to automatically bypass SecureDNA denials of sequences on the lab exemption list as long as they are shipped to the lab's address (Fig. 5d). In all cases, a physical hardware authentication key ($20-$30) or time-based one-time password (TOTP) from an authentication app is required to validate the user's identity, and placing an order for an exempted controlled sequence notifies the principal investigator, the biosafety officer, and the organization's legal contact.

For example, suppose that a laboratory wants to develop a broad-spectrum influenza vaccine and needs access to 1918 influenza. Using the exemption request tool, they can select "1918 influenza," then send the request to their biosafety authority (Extended Data Fig. 3). Once approved, every lab member can use the ET and their personal certificate proving lab affiliation to ship any DNA sequences corresponding to that strain to the lab's registered address, or synthesize them on a benchtop device. Researchers can also request exemptions covering lists of sequences, such as oligonucleotide libraries for deep mutational scanning[21] and directed evolution[22]. Importantly, the database tracks which pieces of controlled sequences have been synthesized by any provider, with or without exemptions, in order to detect split-order attacks.

As a final precaution, whenever an order for a public controlled sequence is automatically approved using an exemption list, the SecureDNA database notifies the lab's principal investigator, biosafety authority, and institutional legal department for record-keeping purposes. Because biosafety vetting and approval are available as a commercial service[23], researchers anywhere in the world can access the benefits of seamless and secure ordering via automated customer screening.

## Discussion

Progress in the life sciences is demonstrably vulnerable to public mistrust[24]. A pandemic deliberately caused by a malicious actor would almost certainly trigger a backlash and draconian research restrictions. Safeguarding the promise of biotechnology requires a way to ensure controlled DNA is only shipped to legitimate laboratories[25,26], but the complexity of the problem and the expense of regulation have deterred international action.

The advent of free, automated, benchtop-compatible, and privacy-preserving DNA synthesis screening (Extended Data Table 2) may allow nations currently hesitant to place their own companies at a competitive disadvantage to begin regulating the sector and clarify liability in the event of misuse. Indeed, signatories of the Biological Weapons Convention may be obligated to require free screening under Article IV, which states that parties must "take any necessary measures to prohibit and prevent the development, production, stockpiling, acquisition, or retention of the agents, toxins, weapons, equipment and means of delivery... within the territory of such State, under its jurisdiction or under its control anywhere."

While DNA synthesis screening may prevent widespread access to credible pandemic agents for many years, advances in *de novo* protein design[27–29] will gradually undermine its effectiveness. Design models can already generate functional equivalents of binding proteins; while this presumably includes toxins, they are far less relevant to international security than pandemic agents. However, *de novo* biodesign tools may eventually be capable of generating allosteric or catalytic proteins sufficient to enhance natural pathogens or even produce novel pandemic agents. Function and folding prediction tools that rely on a complete and intact sequence will not be able to detect such threats if the resulting synthetic genes are ordered in pieces, among other evasive strategies (Extended Data Fig. 4).

Controlling access to protein and genome design tools via APIs and logging controlled designs for screening purposes, which could be done for nearly all users even if source code was shared among all registered developers, may help prevent redesigned versions of controlled agents from evading detection. To further mitigate security risks, leaders in the protein design community have called for the retention of cryptographic records of DNA synthesis orders[30], which could deter malicious actors by reliably identifying the source of the harmful DNA after the fact. Systems running SecureDNA could readily store order hash records, with old keys divided among trusted authorities after rotations and released only following a catastrophic event, to enable forensic screening of past synthesis activity while maintaining day-to-day privacy[30].

Collectively, our results suggest that SecureDNA can provide free, private, reliable, and verifiably up-to-date nucleic acid synthesis screening at a scale sufficient to meet global demand, with negligible false alarms and a largely automated authorization process. If supported by favorable regulatory and liability policies, hardware integration into next-generation synthesizers, and subsidized trade-in programs, near-universal screening could dramatically reduce unauthorized access to pandemic-class agents without delaying research. By preventing the conversion of controlled blueprints into dangerous pathogens, we can safeguard biotechnology and the world from the threat of deliberate pandemics.

## Methods

*Software*

The SecureDNA code base is written in Rust, a "memory-safe" language designed to prevent a wide variety of typical programmer errors which account for roughly half of all security issues, such as mishandling storage allocation or exceeding array bounds[31]. All code, except that used for database generation, is available at https://github.com/SecureDNA/. The public demo is hosted at https://securedna.org/demo/.

We initially implemented the system using Amazon Web Services (AWS) as specified, plus one Database Server and two Keyservers using Google Cloud Platform (GCP) with similar specifications:

- Synthesizer: one `c5d.2xlarge` instance (8 CPU threads at 3.0 GHz; 16 GiB memory)
- Database Server: one `c4.2xlarge` instance (8 CPU threads at 2.9 GHz; 16 GiB memory; gp3 disk)
- Keyservers: three `t3.xlarge` instances (4 CPU threads at 2.5 GHz; 16 GiB memory)

To demonstrate high-volume/high-rate screening (e.g., 40Mbp in under 10 minutes), the client and keyservers each used `c5a.16xlarge` instances (64 threads), and a physical database server (AMD Ryzen 9 5950X, 32 threads, with Samsung 980 Pro NVMe; less than $2K purchase price for entire machine). Because a `c5a.16xlarge` costs USD$2.464/hr on-demand, the marginal cost for a client in the cloud to screen 1 million bp is 1 cent.

To perform screening on a DNA sequence (FASTA) file, `synthclient` opens network streams to (at least) a threshold number of keyservers and a single database. It calls `quickdna` to generate subsequences for each window of the relevant size, each of which is obliviously hashed with the keyservers; unblinds the output; and sends the resulting hash to the database, which responds "Yes," "Yes, but" (allowed via exemption token), or "No." For a public threat (currently all database entries), a denial is accompanied by index information, allowing the frontend of `synthclient` to present a visual analysis for the user.

*Database generation*

To create the version of the database used for testing, listed U.S. Select Agent, Australia Group, EU, and Chinese pathogens were given region tags and separated based on whether they came from viruses, toxins, or microbes. All 30-mer, 42-mer, and 60-mer windows from viruses, toxins, and genes encoding toxins or capable of conferring pathogenicity upon a more easily obtained avirulent microbe (e.g. the three toxin and five capsular genes of *B. anthracis*) were extracted and the 60-mers translated into peptides. All single and double mutants of 42-mers were included along with a number of additional 30-mer and 42-mer predicted functional variants[19]. Peptide windows were selected quasi-randomly and functional variants predicted using a combination of fuNTRp and BLOSUM62. For genes from regulated pathogens that cannot confer pathogenicity, one 42-mer for every 39-45 nucleotides was included and tagged "Regulated but Pass" to identify unregulated sequences from controlled organisms. Controlled subsequences commonly used in biotechnology were tagged as "Common" and were not used to generate variants or trigger denials during screening. Finally, all subsequences with Shannon entropy below 1.6 bits, often found in many unrelated organisms, were removed.

*Database curation*

Non-redundant nucleotide (nr/nt) and protein databases were downloaded from NCBI and subjected to taxonomic and keyword analysis to detect relatedness to controlled genes and functions. The remaining unregulated sequences were separated into 30-mer, 42-mer, and 20-aa windows grouped by accession number (AN). Candidate subsequences from controlled agents were compared to the database of unregulated sequences. Matches to unregulated ANs that exceeded a threshold number of matches to a single candidate controlled sequence were ignored as too closely related. For all remaining matches, the responsible subsequence was removed from the database to minimize false alarms.

*Performance*

To determine server provisioning requirements, nucleotides screened per second (nt/s) were quantified for each component by issuing randomized screening or cryptographic protocol requests as appropriate. A proprietary dataset of over 42,000 real customer orders, the combination of data shared by three different providers, was screened using the SecureDNA alpha prototype to assess performance under ideal conditions on realistic production traffic. To model benchtop synthesis conditions, wall-clock and user mode execution times were measured on a Raspberry Pi 4 client connected via WiFi screening random DNA sequences from 100 to 100,000 bp in length. The Raspberry Pi demonstrated performance exceeding 100 bp/s per thread, indicating feasibility for low-cost embedded CPUs (Extended Data Table 1). In addition, we have demonstrated that large oligo providers can screen 40Mbp orders in under 10 minutes for less than one dollar. Together, these complementary tests quantified end-to-end latency, server capacity needs, and embedded processor utilization relevant to global deployment.

*Sensitivity challenges*

Three proprietary datasets generated by DNA synthesis providers were used for challenge testing. A positive control dataset including sequences from all species on the United States, EU, and Australia Group control lists was used to verify that all controlled sequences were detected. A second test set included controlled and non-controlled sequences lightly manipulated for obfuscation[16]. The final dataset included interspersed subsequences of controlled and non-controlled sequences to test sensitivity for *k*-mers of different lengths. These tests were performed in addition to those described in the companion paper[19].

*Specificity testing*

Anonymized proprietary sequences from commercially synthesized genes were cryptographically screened via SecureDNA without disclosing confidential information (SI Appendix E). Results were conveyed to industry partners for analysis. Analyses of each individual gene flagged or denied by SecureDNA were conducted via nucleotide and/or protein BLAST, with as-yet-unidentified regions subjected to iterative analyses until all subsequences were identified. The SecureDNA results were then scored to quantify specificity.

*Exemption tokens*

Exemption tokens (ETs) are cryptographically-signed objects containing the exemption signed by the last certificate in a chain extending to the root held by the SecureDNA Foundation (Extended Data Fig. 2). Requests are generated using the browser-accessible `ETR` tool, and can be approved by biosafety authorities and ETs issued using the `Exemption System GUI` tool.

## Acknowledgements

We are deeply grateful to several anonymous industry partners, and to B. Mueller for coordination.

## Author Contributions

A.C.Y., D.G., K.M.E., L.F., and R.R. conceived the study; Y.Y., C.B., and M.G. wrote the cryptographic analysis with advice from A.C.Y., I.D., R.R., D.W., V.V., and A.S.; D.G. and K.M.E. created the figures, A.C.Y., M.G., Y.Y., C.B., I.D., K.L, J.L., X.L., Y.W., and H.Z. contributed to system design; software systems architecture & security was led by L.F.; the software implementation was overseen by L.F. and J.B. and performed by L.F., J.B., T.V., M.K., W.C., F.S-L., L.V.H., S.W., and B.W-R.; database curation was led by T.V. and L.F. with input from K.M.E. and D.G.; performance assessments were performed by L.F. and D.G.; specificity analyses were conducted by D.G., L.F., T.V., and K.M.E.; and the certificate design and implementation were led by L.F. and F.S-L. All authors contributed to writing the paper.

## Funding

Financial support was obtained from the Open Philanthropy Project (to MIT, Aarhus University, and SecureBio), an anonymous philanthropist from mainland China (to Tsinghua University), the Aphorism Foundation (to MIT and SecureBio), and Effective Giving (to MIT and SecureBio). The funders had no role in study design, data collection, data analysis, data interpretation, or writing of the report. To preserve international neutrality, no government funds were used to support this project.

## Competing interests

K.M.E. and D.G. are authors of PCT/US2021/014814 filed by the Massachusetts Institute of Technology. All authors share an interest in preventing future pandemics.

## Data availability

The privacy of customer data from historical DNA synthesis orders is protected by legal agreements signed by the relevant providers. We encourage interested parties to reach out to providers directly to discuss the possibility of signing non-disclosure agreements.

## Code availability

The open-source SecureDNA software is available at https://github.com/SecureDNA/.

Correspondence and requests for materials should be addressed to esvelt@mit.edu, to ivan@cs.au.dk, or to andrewcyao@tsinghua.edu.cn.

**a**

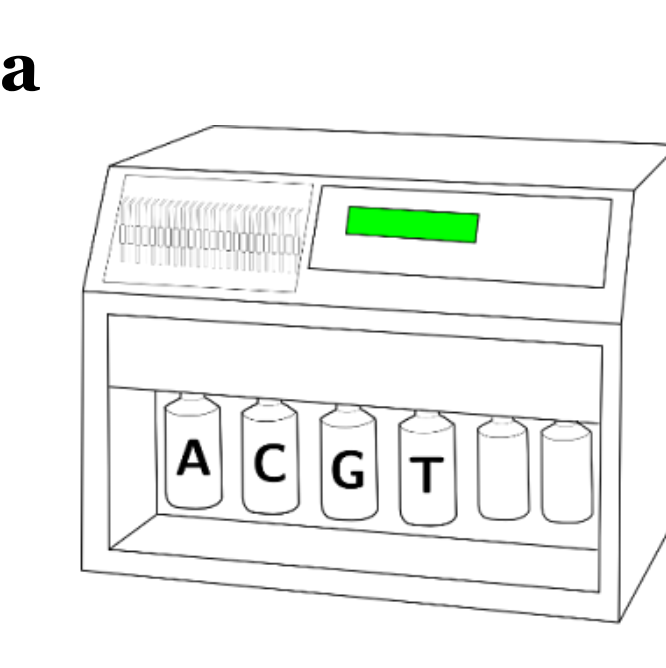

**Hazardous sequence**

GGTACCAGCTATTCACGACTGTC...

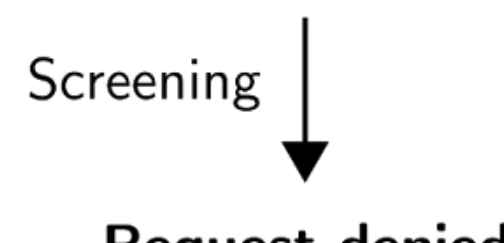

**Request denied**

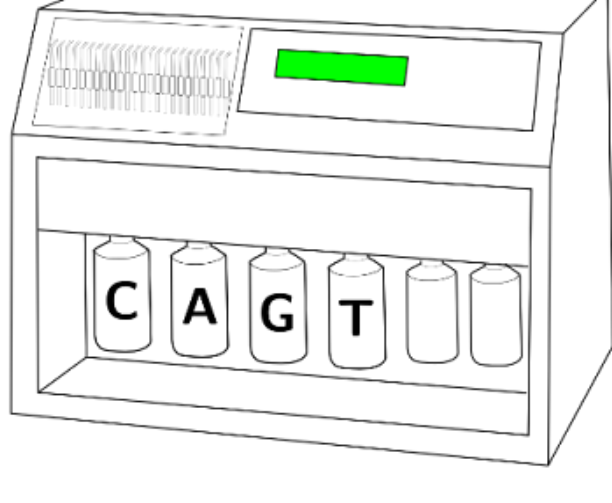

**Permuted sequence**

GGTCAACGATCTTACAGCATGTA...

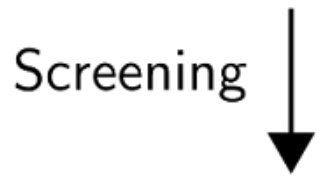

**Hazard synthesized**

GGTACCAGCTATTCACGACTGTC...

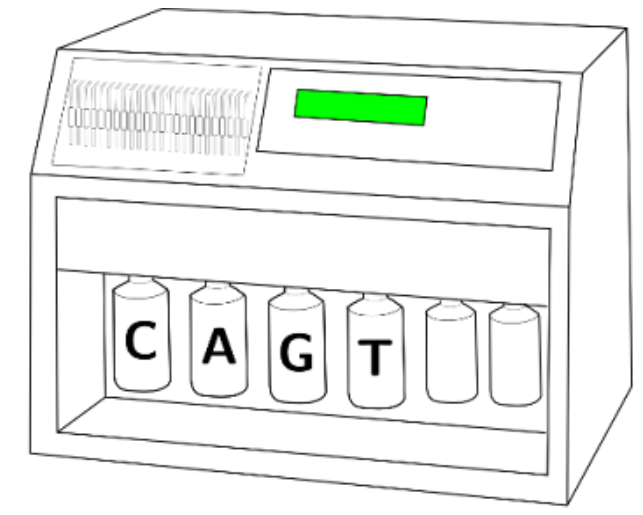

**Permuted sequence**

GGTCAACGATCTTACAGCATGTA...

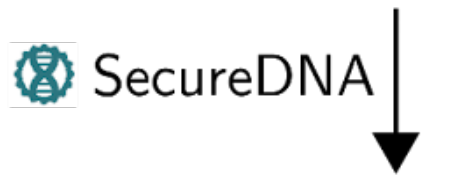

**Request denied**

**b**

| Permutation 1 | Permutation 2 | Permutation 3 | Permutation 4 |
|---|---|---|---|
| A ↔ C | - | - | - |
| A ↔ G | - | - | - |
| A ↔ T | - | - | - |
| C ↔ G | - | - | - |
| C ↔ T | - | - | - |
| G ↔ T | - | - | - |
| A ↔ C | G ↔ T | - | - |
| A ↔ G | C ↔ T | - | - |
| A ↔ T | C ↔ G | - | - |
| A ↔ C | C ↔ G | G ↔ A | - |
| A ↔ C | C ↔ T | T ↔ A | - |
| A ↔ G | G ↔ C | C ↔ A | - |
| A ↔ G | G ↔ T | T ↔ A | - |
| A ↔ T | T ↔ C | C ↔ A | - |
| A ↔ T | T ↔ G | G ↔ A | - |
| C ↔ G | G ↔ T | T ↔ C | - |
| C ↔ T | T ↔ A | A ↔ C | - |
| C ↔ T | T ↔ G | G ↔ C | - |
| A ↔ C | C ↔ G | G ↔ T | T ↔ A |
| A ↔ C | C ↔ T | T ↔ G | G ↔ A |
| A ↔ G | G ↔ C | C ↔ T | T ↔ A |
| A ↔ G | G ↔ T | T ↔ C | C ↔ A |
| A ↔ T | T ↔ C | C ↔ G | G ↔ A |
| A ↔ T | T ↔ G | G ↔ C | C ↔ A |

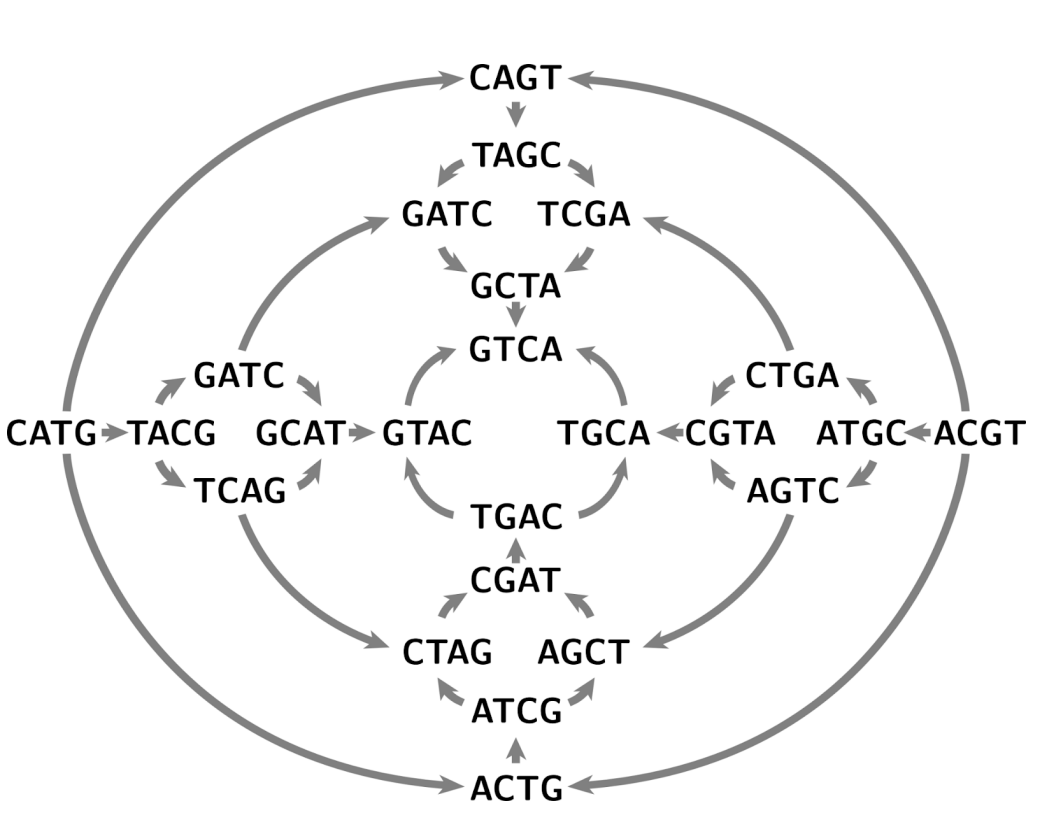

**Extended Data Fig. 1 | Permutation attacks on benchtops. a)** Many DNA synthesizers use four distinct reagents to add each of the four bases. Anyone with access to the machine can swap reagent bottles and permute the affected bases in their order to obtain the same DNA sequence. **b)** SecureDNA can screen benchtop queries for all 24 possible permutations of each subsequence to prevent reagent manipulation, using a technique in which all permutations, in both the forward and reverse-complement directions, are mapped into a single hash. Consequently, defending against such swaps incurs no performance cost.

## a. Cloud Deployment

| Thread Speed | Cost per 40Mbp screened | Cost per 1Mbp |
|---|---|---|
| 1275 bp/thread/sec | ~$0.40 (1 cent) – high-end CPU | 1 cent |
| 1275 bp/thread/sec | ~$0.20 (½ cent) – mid-end CPU | ½ cent |
| 1275 bp/thread/sec | ~$0.10 (¼ cent) – low-end CPU | ¼ cent |

| Instance Type | Hourly Cost | Performance |
|---|---|---|
| 64-thread c5a.16xlarge | $2.4640/hr | 40Mbp in 9m (40 cents) |
| 1-thread t2.small | $0.0230/hr | 1Mbp in 15m (0.5 cents) |
| 1-thread t2.micro | $0.0116/hr | 70Kbp in 1m (0.02 cents) |

## b. Local Deployment

| Hardware | Capital Investment | Performance |
|---|---|---|
| AMD ThreadRipper | ~$4000 | 40Mbp in <10m |
| AMD Ryzen 9 5950X | ~$2000 | 20Mbp in <10m |
| Raspberry Pi 4 | ~$50 | 42Kbp in <1m |

**Extended Data Table 1 | Client cost comparisons for cloud deployment and locally-owned hardware.**
**a)** Cloud deployment is roughly a fixed cost per base pair (bp), with capital investment = $0. It assumes servers are shut down when not screening and that costs when off, e.g., fixed IP addresses, are negligible.
**b)** Local deployment has a fixed hardware cost, but the marginal cost of screening is effectively $0 since power demand during screening is negligible. At $0.10/kWh and ~200W TDP for a high-end CPU, 10 minutes of screening costs approximately 0.3 cents. All figures are in USD.

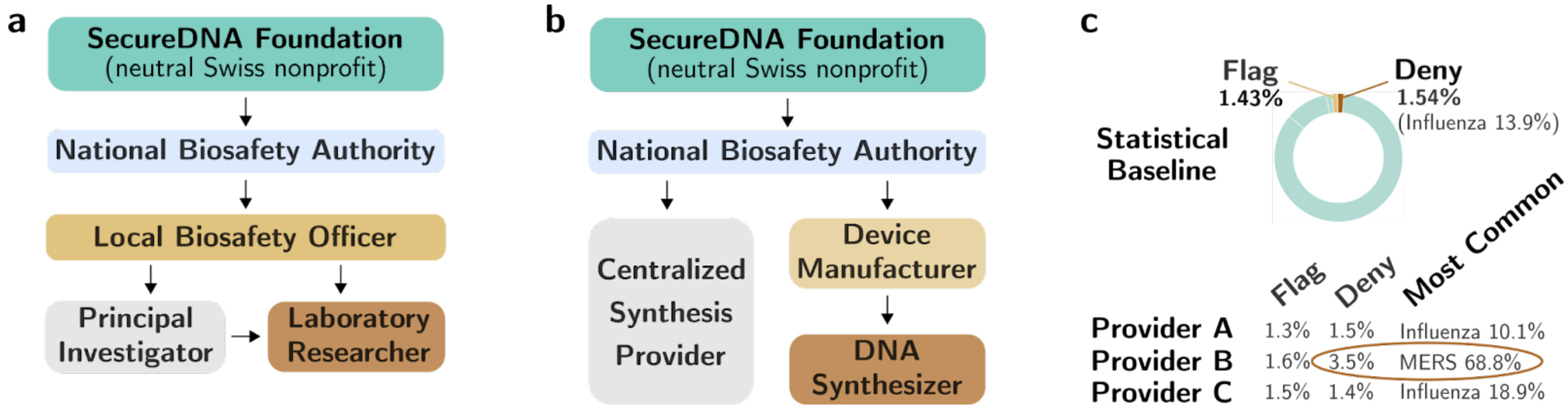

**Extended Data Fig. 2 | SecureDNA certificate chains. a)** The Switzerland-based SecureDNA Foundation issues exemption certificates to each national biosafety authority, which in turn can issue certificates to local biosafety officers, which can issue certificates to principal investigators, and then to laboratory researchers. Researchers can either use a one-time exemption token issued by their local biosafety officer or a laboratory exemption token together with the certificate issued to them by the lab's principal investigator. **b)** The SecureDNA Foundation issues screening certificates to each DNA synthesis provider and manufacturer of DNA synthesis machines. Each machine receives a certificate, which accompanies every screening order. **c)** The SecureDNA system records the number of matches to different public controlled sequences associated with each certificate for analysis. Statistical associations concerning the number and pattern of matches to the database can detect anomalous adversarial activity indicative of dictionary or other attacks.

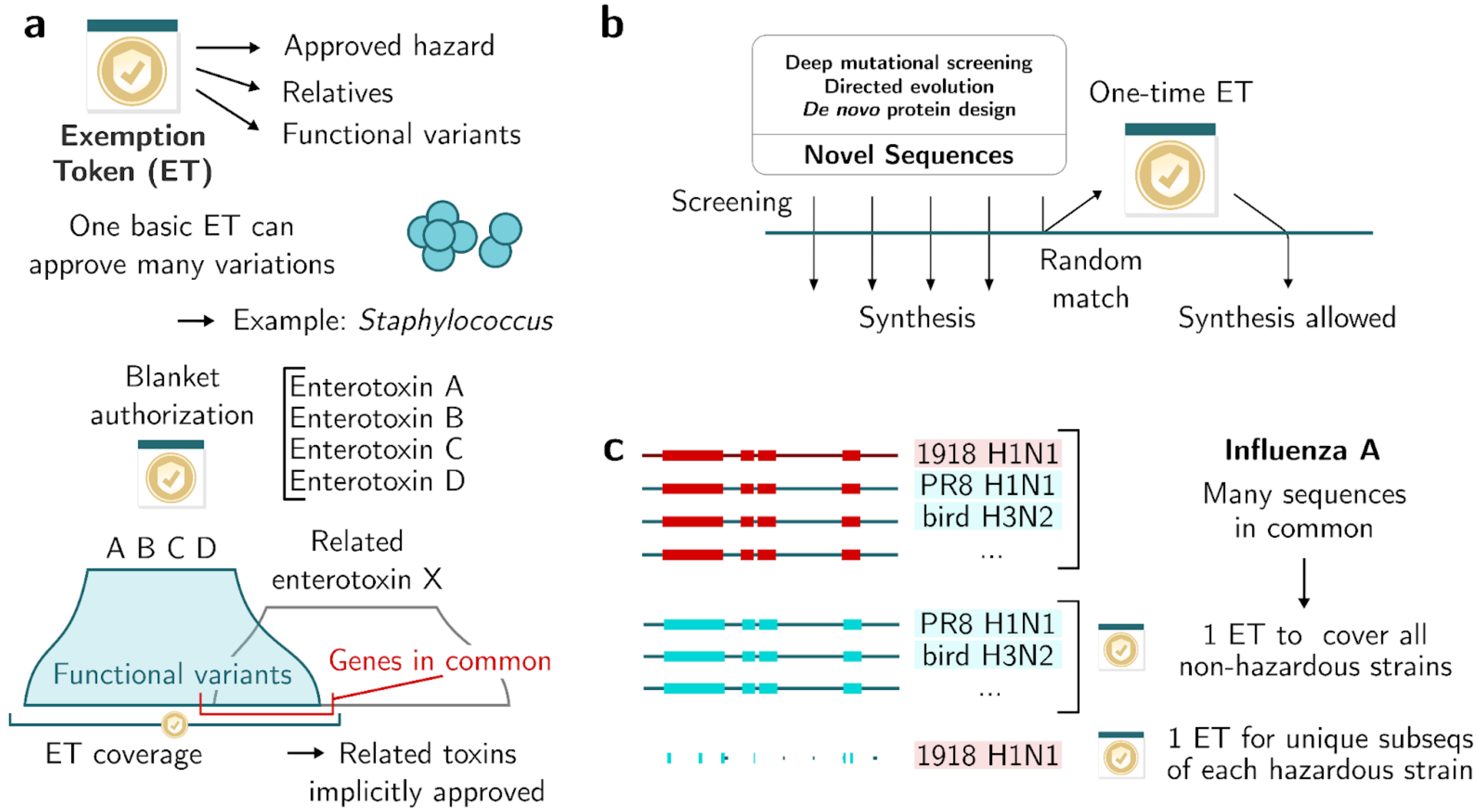

**Extended Data Fig. 3 | Exemption token versatility. a)** A laboratory exemption token (ET) allows members to obtain any controlled sequence approved by their local biosafety authority, as well as close relatives and predicted functional variants with k-mers shared with the controlled sequence. For example, a laboratory that works with staphyloccocal enterotoxins will receive an ET linked to the primary accession numbers (ANs) of the relevant toxin-encoding genes of subtypes A through E. Because closely related toxins with different ANs would normally be recognized during screening by the subsequences they share with the primary ANs and predicted functional variants, the ET gives access to all such toxins in the group. To avoid accidentally granting too much access due to oversight or misunderstanding, requests for organism-wide exemptions prompt an additional warning to the biosafety officer. Specifically, the system displays the message: *"If granted, the laboratory will have access to all relevant genes from X. This should not be granted unless they are authorized to work with the intact organism."* **b)** Oligonucleotide libraries for deep mutational scanning or directed evolution experiments and *de novo* designed genes do not correspond to sequences in repositories and may match a random controlled subsequence. If this occurs, researchers can request a one-time exemption list token that will pass the specific set of sequences requested; because it is rare, unregulated orders should not expect many such matches. **c)** The system adequately distinguishes between 1918 H1N1 influenza and other variants, and other controlled influenza strains with closer relatives such as high-pathogenicity avian influenza are defined by phenotypic assays rather than genome sequence. If this changes, and a controlled influenza strain has too few unique k-mers to adequately defend, then additional k-mers shared with other strains might be included to trade specificity for sensitivity, but the shared and unique k-mers might be differentiated using exemption tokens. For example, a "shared influenza A" ET covering the shared sequences might be required, with additional ETs required to access unique sequences from controlled strains (as is currently the case).

| | **SecureDNA** | **Current alignment screening** |
|---|---|---|
| Reliably detects publicly recognized regulated pathogens | Yes | Yes |
| Speed (asymptotic, Big O notation) | $O(s)$ (hash-based methods enable constant time per window independent of database size) | BLAST: $O(s \cdot d)$ worst-case, $O(s + d)$ in practice (since $d \gg s$, run time increases with increasing database size) |
| Preserves customer and provider privacy | Yes, cryptographically secure | No, preventing alignment algorithms from learning input sequences would require homomorphic encryption too inefficient to use at scale[32] |
| Minimum window size | ≥ 30 nucleotides or 20 amino acids | ≥ 200 nucleotides (typical) |
| False alarm rate | Curation to remove unregulated matches<br>→ no nonrandom false alarms | Many matches to unrelated genes require human review |
| Fully automatable | Yes, can provide baseline denial of all regulated sequences in hardware | No, requires human review for all matches |
| Compatible with benchtop synthesizers/assemblers | Yes, given a secure connection | No, benchtop users can ignore screening |
| Resistant to evasion by mutation or algorithmic design | Yes, these techniques are applied when building the database, and the database is private | No, public nature of screening tool allows adversaries to fine-tune sequences that pass screening |
| Can screen for emerging threats without disclosure | Yes[33] | No, algorithm matches plaintext database, which requires disclosure to screener; too inefficient to encrypt at scale[32] |
| Can run locally with no Internet connection | No, requires connection to stay up-to-date on new regulations and emerging threats, and must be remote to prevent evasion by fine-tuning or via split orders | Yes, but cannot stay up to date and cannot resist fine-tuning from repeated queries |

**Definition of variables**
Order size (nucleotides): $s$
Database size (hash entries): $d$

**Extended Data Table 2 | Characteristics of SecureDNA compared to current alignment-based fuzzy screening approaches.**

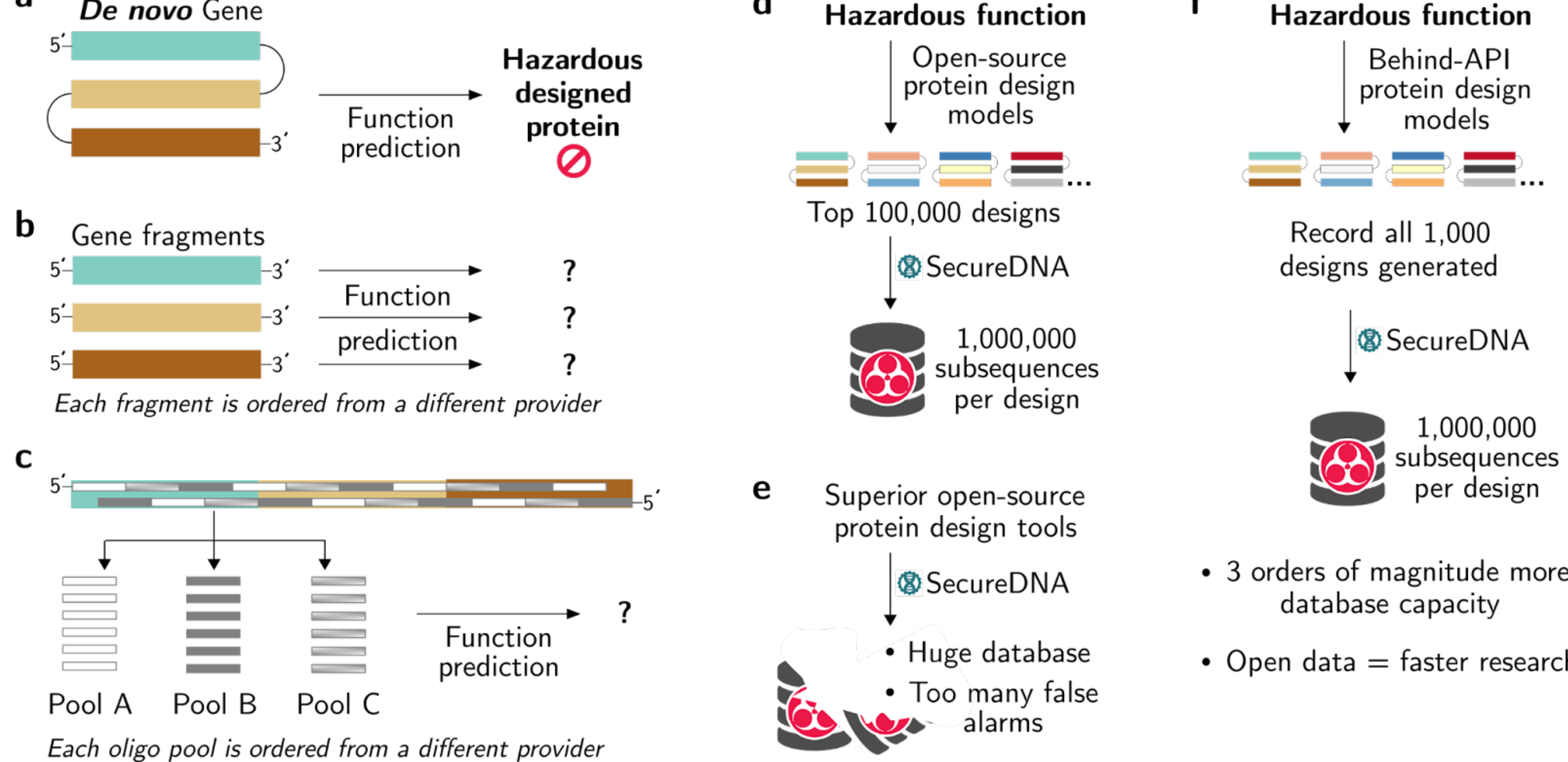

**Extended Data Fig. 4 | *De novo* designed proteins will eventually evade screening. a)** Protein design tools will become increasingly capable of generating sequences with desired functions, including those of controlled agents. In principle, the same tool could predict the function of the generated sequence. **b)** In practice, these tools cannot be used for screening because designs can be generated in pieces with unpredictable folding and activity patterns. Dividing a designed sequence into three parts will generate three distinct polypeptide chains that are unlikely to fold into any structure with a predictable function. Ordering them from different providers at different times will preclude function prediction. **c)** Similarly, designed sequences can be assembled from pools of single-stranded oligonucleotides ordered from three separate providers, ensuring that no single provider can access the complete sequence, let alone predict its function. **d)** In principle, SecureDNA could defend most subsequences common to the top predicted controlled sequences from leading public design tools. Since only a small fraction of DNA synthesis orders risk random false alarms because they are not present in repositories, it might be possible to generate a controlled sequence database with as many as $10^{14}$ database entries while triggering only one random false alarm per month. Such a database could theoretically dedicate a million subsequences to defending each of 100,000 *de novo* designs for each of a thousand controlled functions, although practical limitations may arise. **e)** Even given peptide screening, it will still be possible to generate designed hazards by assembling oligonucleotides too short for peptide screening, and the false alarm rate will gradually increase due to the need to defend more sequences as tools improve. Screening will eventually fail once enough *de novo* designs are possible for a given function. **f)** If protein design tools are only available through an API, and all designed sequences are logged, potentially hazardous designs can be included in the database and reliably detected.

**Supplementary Information**

**“A system capable of verifiably and privately screening global DNA synthesis”**

## Appendix A: Maintaining an up-to-date database

If the database is not kept up-to-date, adversaries keeping abreast of the literature and news will be able to immediately place orders and obtain newly credible pandemic viruses and other biological weapons. This is a major security vulnerability for all distributed solutions to DNA synthesis screening. SecureDNA solves this problem by maintaining a single database of subsequences from currently known controlled agents and updating it every other day.

Specifically, SecureDNA staff maintain automated web alerts for potential pandemic pathogens and biological weapons. When a credible new agent is reported in a preprint or publication, all wild-type 30-mer subsequences are selected, encrypted via the keyservers, and the results added to the database without functional curation to provide immediate protection. Meanwhile, functional variants are predicted and curation is performed to generate subsequences capable of both sensitive and highly specific protection. Once encrypted, these new database entries replace the stopgap entries generated from the uncurated wild-type 30-mers. Systems in which government-run sequence databases notify the screening system before posting a wild-type sequence are possible. Once future red-teaming and prize competitions comprehensively assess the integrity of the SecureDNA architecture and nations discuss the feasibility of addressing emerging hazards, it may be possible to establish a system to add threats that are not yet publicly credible to the database without disclosing their identities to anyone beyond the concerned researcher who flags the hazard and a single approved expert with database addition access.

Similar alerts monitor the addition of novel threats to government lists of controlled agents. When this occurs, the entries associated with the affected agent are updated with suitable region tags to ensure compliance with local regulations and export control laws. If periodic literature reviews uncover evidence that an unregulated threat is no longer credible, the corresponding entries can be removed from the database.

## Appendix B: Analysis of information leakage and vulnerabilities

The SecureDNA system deliberately discloses small amounts of information to various parties as part of its normal operation. In addition, depending on the threat model, it could also leak some information to various types of attackers. However, under realistic assumptions about how actual synthesis customers order DNA, these do not significantly compromise customer privacy. The following sections document such information flows, plausible threat models, and detail why full private-set intersection methods are both unworkable and undesirable.

### Typical information flows with no attacker

Under normal circumstances (e.g., absent an active attacker), the chart below documents what information is routinely available to various parties in the system.

- **Parties:**
    - **C** customer
    - **O** BSO
    - **P** provider
    - **B** benchtop
    - **V** benchtop's vendor
    - **S** SecureDNA
- **Information:**
    - **s** actual DNA sequence being ordered
    - **e** contents of a particular exemption for C
    - **p** provider identity (company name, domain)
    - **b** benchtop serial number
    - **v** benchtop vendor identity (company name, domain)
    - **c** customer identity (name, email, shipping address)
    - **o** BSO identity (name, email, institution)

### Order without exemption, whether granted or denied (thus no controlled DNA was produced)

| Party→<br>Info↓ | C | O | P | B | V | S | Notes |
|---|---|---|---|---|---|---|---|
| s | y | - | y | y | n | n[5] | Both the customer and the entity making the DNA must possess this information |
| e | - | - | - | - | - | - | No exemption is present for this order |
| p | y | - | y | - | - | y | The customer knows the provider with whom they've placed the order |
| b | y | - | - | y | y | y | The customer may know the serial number but typically has no operational need for this information |

| **v** | y | - | - | y | y | y | The device manufacturer is likely known to the customer but may not be operationally relevant |
|---|---|---|---|---|---|---|---|
| **c** | y | - | y | y | n | n | The provider requires customer information for payment and delivery; the device vendor only knows the purchaser of the device |
| **o** | - | - | - | - | - | - | No exemption is involved in this transaction |

### Order with exemption, and in which synthesis of a controlled sequence was granted

(*Italics* indicate a change from the no-exemption case)

| **Party→ Info↓** | **C** | **O** | **P** | **B** | **V** | **S** | **Notes** |
|---|---|---|---|---|---|---|---|
| **s** | y | *y*[1] | y | y | n | n[5] | |
| **e** | *y* | *y* | *y* | *y* | *y*[2]/n | y[3] | All parties who handle the exemption know *which controlled sequences* it covers |
| **p** | y | *n* | y | - | - | y | |
| **b** | y | *n* | - | y | y | y | |
| **v** | y | *n* | - | y | y | y | |
| **c** | y | *y* | y | y | *y* | *y*[4]/n | |
| **o** | *y* | *y* | *y* | *y* | *y* | *y* | |

### Notes

[1] The BSO knows the *maximal set* of possible sequences ordered by customer, but has no idea whether any order ever used it, which subset is in any order, when they were placed, or how many times, unless the BSO requests in its signing certificate that usage notification be enabled.

[2] If a benchtop vendor has issued a machine certificate which requires audit reporting, then SecureDNA will send encrypted email to the vendor indicating that this customer's exemption covered a controlled sequence.

[3] SecureDNA knows the number of windows in the order which were covered by the exemption, but does not know *what DNA sequence* those windows apply to.

[4] If exemption is not blinded.

[5] When doing split-order detection, then for *only* windows which are (a) regulated, (b) wild-type, and (c) viruses, then SecureDNA could in theory know these because backend infrastructure increments a counter for each one per-order, including which provider/benchtop they came from.

**Active attacks**

For various possible capabilities, what can an active attacker learn?

All of the scenarios presented below assume an active attacker who possesses sufficient technical skill to persistently compromise Internet servers, and who is intent on discovering what particular customers are synthesizing. They also assume that the attacker is only attempting to compromise *SecureDNA*, which is unrealistic for many plausible attacks—many of the possible attacks are likely easier to accomplish by compromising *a provider*, either by direct attack on their infrastructure, or via out-of-band methods such as bribing, blackmailing, or coercing employees instead; or instead by compromising personnel at the customer end, up to and including intercepting a shipment from the provider (e.g., by bribing a mailroom employee, or misleading a delivery person) and then sequencing the pilfered order. We similarly assume that the attacker's goal is *information disclosure* of customer orders (perhaps as a competitor, or in some sort of stock-manipulation scheme), rather than simply making the system unavailable in general (which can be more simply accomplished by attacking the network at either end, e.g., via denial-of-service attacks or physical attacks on network infrastructure). Finally, we assume that passive eavesdropping of communications channels is not possible, since SecureDNA uses TLS on all connections and eavesdropping such connections is equivalent to breaking TLS.

Given the above assumptions, an attacker who is able to establish a persistent presence (such as a root-level compromise of the server) on a single keyserver can learn the (approximate) total *volume* of orders being processed. (This is approximate because not every order goes through every keyserver.) By inspecting the offered synthesis certificates, such an attacker may deduce *which synthesis providers* are using the server, but this does not disclose *which customers' order* are being screened, because data sent to keyservers does not identify the individual customer placing the order, but only the synthesis provider which is performing the screening. If the customer is instead using a benchtop synthesizer, then the benchtop's synthesis certificate identifies *the machine* (e.g., by its serial number), but this does not leak *which customer* the benchtop belongs to, unless the attacker can also compromise the benchtop vendor's sales records. The benchtop's IP address may be a clue, but customers are free to use VPNs to hide their originating IP addresses; this is one of many reasons why SecureDNA does not use IP addresses for any sort of authentication. As for determining which windows are being screened, a single keyserver's data cannot help an attacker because of the use of threshold cryptography.

An attacker who compromises *multiple* keyservers (at least the threshold number of them) still learns nothing about the hashes being sent (much less the actual sequences), because of the cryptographic blinding which is applied by synthclient before data is sent to the keyservers; this blinding is removed again by synthclient after receiving the keyserver responses.

An attacker with persistent access to one or more database servers is in a more powerful position, but is still quite limited in what they can discover. First, unless they have compromised all database servers, they only have a *1/d* chance of intercepting any particular customer order, where *d* is the number of database servers. (Of course, a sufficiently powerful attacker who can also control the customer's networking may also force selection of a particular database by making the rest of them unreachable by the customer, e.g., via a DDoS attack.) However, even if the attacker is able to monitor the use of a database by a particular customer, an attacker who intercepts an order which does not include an exemption certificate and which does not include any controlled sequences gets only the list of window hashes and the resulting database response of "no controlled sequences detected." Because each hash was generated in a cryptographically-secure fashion from the original DNA or AA window, inverting this hash amounts to creating a table of every plausible DNA or AA window which might be generated by the customer, e.g., computing every plausible such sequence, hashing it through the keyservers, then comparing each of those hashes to the incoming hash stream from a given customer. Furthermore, the attacker must do all of this during the same database-rotation generation as that customer's order, since after a rotation, prior hashes are useless. Even though not *every* possible DNA or AA window sequence must be generated (not every possible kmer occurs in nature), the number of window-sized subsequences in NCBI's database is roughly 40 trillion, which is six orders of magnitude larger than any plausible order or set of orders from even the largest single customers. An attacker who has only compromised a database server must thus ask the keyservers to hash all of many guesses, but no customer would have synthesis certificates which allow such a high volume of hashing, so the attacker would very quickly hit rate limits imposed by each keyserver. (An attacker who has compromised the threshold number of keyservers could evade rate limits by stealing keys and performing the hashing elsewhere; if they have also compromised a database server, they could then compare these hashes to the pilfered database traffic. But this violates our threat model, where we assume that no single entity can compromise at least the threshold number of keyservers simultaneously.)

Thus, an attacker has a difficult choice: If they simply wish to *confirm* whether a given customer is synthesizing sequences from a particular organism, they must choose a set of windows unique to that organism, use a pilfered certificate to hash those into a table, then compare their results to those from compromised database server. The more common the organism, the less *useful information* the attacker learns—learning that a customer is synthesizing a sequence used ubiquitously in the industry tells the attacker little. But if the attacker is not trying to confirm a guess of a common organism, or worse yet has no idea what the customer is synthesizing but must instead attempt to survey all possible organisms, the attacker must generate an enormous rainbow table, and doing so will be rate-limited.

If a customer has submitted an exemption certificate, the exemption identifies which sequences the customer may wish to order, and this exemption is in plaintext when handled on the database server. In addition, the certificate also identifies the customer, unless the exemption is blinded (which customers may request if their BSO, and all entities higher in that particular exemption chain, allow it; this is a per-chain administrative decision). This is the most powerful attack that an attacker who can reside inside the database server can exploit, because the attacker can see not only which customer has requested which exemptions, but also which ones were actually used in the order. The SecureDNA system makes it straightforward to create and use minimal exemptions which cover *only* the sequences used in one particular order, hence leaking as little as possible to an attacker who is able to compromise a database server. Since most customers do not order controlled sequences and thus have no need of exemptions, this attack is only of use against customers who order controlled sequences, and nonetheless requires that an attacker has surreptitious access to the database server's decrypted traffic (e.g., have actively compromised the running server itself) at the exact moment the customer's order was screened.

**Avoiding leakage**

One potential mitigation of the threat from the most powerful attacker—one who has complete, persistent control of at least one database server—would be to use cryptographic algorithms such as private information retrieval or private set intersection. This would prevent such an attacker from either monitoring multiple customers and deducing (from identical hashes) whether two customers have submitted any windows in common (whether controlled or not), or from monitoring actual regulated pathogen hits returned by the database to customers (or just learning that hits have occurred). We will now show why both solutions are not applicable to SecureDNA.

**Private Information Retrieval**

In Private Information Retrieval, a server has a large database, while the client retrieves the items at specific indices of the database without the server learning the indices. At a first glance, this could be applicable to the SecureDNA setting: the database server would encode the database of hashes in some form such that the client can look up if its hashes are present, e.g. using Bloom filters. The client will learn this result, while the server learns nothing. This, however, makes the solution too private: some of the information flows described above are deliberate, and this would compromise them.

*Too much privacy?*

Were SecureDNA to implement PIR (in some form), this means that an actual controlled sequence hit in customer-supplied data is indistinguishable by the database from any other window screened by the system. But this means that the database would be unable to inform the customer of what kind of regulated pathogen or controlled sequence is present in the order; this is unacceptable because synthesis providers need to employ additional procedures for such hits (such as checking for export controls or performing additional know-your-customer checks). Any detected hit, which usually covers multiple windows, has to potentially generate metadata identifying the organism in case the customer has an exemption for an organism that does not specify predicted functional variants. Moreover, this metadata may sometimes not say which window generated a hit (for certain emerging threats). This application requirement, namely that detected hits must be computed upon to generate specific metadata about the regulated pathogen or controlled sequence, means the metadata cannot be incorporated directly into a purely oblivious PIR response.

In addition, detecting split orders would be defeated in a pure-PIR implementation, in which the central server can know nothing about a customer's order, even if the customer is ordering Select Agents or similar regulated organisms. This would allow a malicious customer to assemble a dangerous organism by ordering many smaller sub-pieces, each below any given vendor's threat threshold, and reassembling them in the lab. In addition, benchtop vendors may be legally required to be able to audit controlled sequences their customers are producing *without* necessarily depending on access to the customer's equipment, and have specifically requested this capability. Further, many lab principal investigators and biosafety officers require visibility into how exemption certificates they have granted are being used; this, too, would be defeated by a pure-PIR implementation. (Note that asking the customer endpoint equipment to issue such reports instead is too easily subverted, e,g., by the customer interrupting such communication, given that the equipment making the report is under customer control.)

**Private set intersection (PSI)**

Another solution would be to use so-called Private Set intersection. Private Set Intersection is a cryptographic tool which allows two (or more) computers, each of which having a list of items, to learn the items that they commonly have on both (or all of their) lists. Clearly if the database server and the provider each input their list of hashes into a PSI protocol which would deliver the output to the database server then this would be applicable to our use case. In our setting, where the input of the provider is much smaller than the database size, one could apply a so-called asymmetric PSI protocol which is more efficient in this case.

*Performance*

The SecureDNA system is useless if it cannot meet customer performance demands. Thus, any PSI solution must scale to the size of our database, which contains around $2^{34}$ elements, in order to properly represent all of the controlled sequences it must detect. In addition, it must be fast enough. Synthesis customers' requirements are bounded at one end by customers who require orders of a few genes (10K base pairs or smaller in total) to be screened in less than ten seconds, and on the other hand by large oligo providers, who routinely run parallel synthesis arrays which require screening 40 million base pairs in less than 10 minutes. Note that this latter rate requires almost 70,000 bp/sec of screening performance, which corresponds to approximately 280,000 hash lookups per second. (Ignoring edge effects from all but the shortest oligos, each bp requires 4 hashes to be looked up—two corresponding to the two different sizes of DNA windows currently in use [our combinational hashing strategy means that one lookup can evaluate all 4!=24 combinations of potentially-reagent-swapped DNA, in both the forward and reverse-complement directions, simultaneously], and two corresponding to both directions and all three reading frames for peptides, divided by three because it takes three bp to make one amino acid.) We have demonstrated this level of performance (74 kbp/sec) in the production system.

*Symmetric PSI*

We consider PSI protocols which are secure against so-called semi-honest adversaries. This matches our existing threat model, and such protocols are also the most efficient. The state of the art in symmetric PSI (where the set sizes of both parties running PSI are about equal) is due to Raghuraman & Rindal [ACM CCS 2022]. Their construction runs in approximately 7 seconds for input sets of size $2^{24}$ (the largest setting they evaluate). The increase in computation time from set size $2^{20}$ to set size $2^{24}$ (the largest they tested) of their construction is a factor 20, so runtime growth to input set size $2^{32}$ (which is still well below what SecureDNA uses) would be at least a factor of 400 assuming an unrealistically optimistic linear growth in runtime of their solution when increasing database size. As SecureDNA must be able to give responses within seconds for input queries of only a few thousand elements, this solution cannot be used.

*Asymmetric PSI*

In synthesis screening, the customer always has a much shorter list of hashes as their input set than the size of the database. This setting is called asymmetric PSI, for which optimized protocols exist. We now consider the state-of-the-art and mention why it does not apply to SecureDNA. Hetz, Scheider & Weinert [ESORICS 2023] (based on an earlier work due to Kales et al. [USENIX Security 2019]) built a protocol that combines an OPRF protocol with a PIR, so the aforementioned problem that PIR only reveals the output to the client occurs again (and is inherent to their approach). Another idea is to use homomorphic encryption, where the client encrypts its inputs and sends them to the server, who compares them homomorphically with its database before sending the response to the client.

Putting that aside, [ESORICS 2023] claim $2^{10}$ lookups in 2 seconds once the precomputation step has been completed—which corresponds to 500 lookups per second—on a cellphone whose specifications compare quite favorably to a mid-to-high-end desktop CPU. This is, alas, more than 500 times too slow to meet our performance demands, given that only one or two such CPUs are easily fast enough to screen 74Kbp/sec at the client end without PSI, and it is unreasonable (and much too expensive) to expect such customers to invest in a multi-hundred-CPU server farm, whether locally or in the cloud, just to enable PSI. Thus, even using asymmetric PSI would require a dramatic, and expensive, expansion of the available computational resources at both ends of the screening process in order to meet the required performance criteria.

Another interesting line of work is the protocol of Demmler et al. [PoPETs 2018]. They avoid the PIR-related fundamental design problem of [ESORICS 2023] that occurs in our setting due to the use of an *inner* PSI which could potentially be modified so that the server obtains the output. However, their construction necessarily requires multiple servers on the backend that are non-colluding, which seems difficult to realize in our setting. We use such an assumption for the key servers, as these perform a very simple task independently and without keeping much state. The different servers in [PoPETs 2018] must hold the same database, and this leads to problems of synchronization and simply having to protect the same secret database on more servers running more complex software.

Kong et al. [ACM CCS 2021] build an OPRF-based solution, and their large preprocessing time (spent on evaluating an OPRF) could potentially be avoided in our setting. Their fast online time for input sets of size $2^{28}$ (<3 minutes) would however, when considering their runtime growth in Table 1 of their work, not be sufficient for a database table size as we have to process. Moreover, they mention that their already very computationally powerful benchmark machine would have to be upgraded to more RAM to support such large databases, making it not applicable in our setting. Finally, their protocol would have to be modified so that the ciphertexts are decrypted to the server, which we believe however is possible. We conclude that while the approach of [ACM CCS 2021] might in the future meet our requirements for SecureDNA, significant performance improvements have to be done before this is possible. The recent work of Mahdavi et al. [USENIX Security 2024] follows a similar approach as [ACM CCS 2021]. Their runtime even for much smaller database sets than the $2^{34}$ items that we require and the client input sizes up to $2^{20}$ would not allow us to respond quickly enough for our customers as outlined above.

**"A system capable of verifiably and privately screening global DNA synthesis"**

## Appendix C: A detailed description of the cryptography underlying SecureDNA

We now describe the cryptography behind the SecureDNA system using standard cryptographic terminology, although only on a high level. This description is aimed at a technical audience. Comprehensive technical details are available in separately published conference proceedings[34].

*Overview*

We consider the SecureDNA system in terms of different entities participating in the system. The three main entities are (1) the synthesizer; (2) the database server; and (3) the curator responsible for populating the database[1]. The synthesizer has inputs $S = \{s_1, \ldots, s_I\}$ which each are bit-strings of arbitrary length. The curator chooses a database $D = \{d_1, \ldots, d_J\}$, also consisting of bit-strings of arbitrary length. Finally, the database server obtains a special version of $D$, denoted as $H$. We require that the leakage of $H$ about $D$ must be kept to a minimum. The goal of SecureDNA is two-fold: we want the synthesizer and database server to engage in an interaction, at the end of which both learn $|S \cap D|$ without either learning any further information about $D$ or $S$. Additionally, we require a protocol which allows the curator to generate $H$ from $D$. To realize these goals, we assume additional entities participating in the protocol, namely $n$ so-called keyservers. On a high level, the centerpiece of our solution is a so-called Distributed Oblivious Pseudorandom Function (DOPRF) [NPR99] where all $n$ keyservers hold a share of a key $k$ of a cryptographic hash function.

*Notation*

We write $Z_p$ for the set $\{0, \ldots, p-1\}$ of residues of the integers $Z$ modulo the prime $p$. The set of bit strings of arbitrary length is denoted as $\{0,1\}^*$. We assume that $G$ is a finite abelian group of order $p$. $G$ is considered in multiplicative notation and we write $\cdot : G \times G \to G$ to denote the group operation. For example, for any $g \in G$ we denote by $g^2$ the value $g \cdot g$ obtained from applying the group operation of $G$ on $g$. We assume that the so-called Decision Diffie-Hellman [Boneh98] problem holds in the group $G$. In practice one can instantiate $G$ e.g. using Groups over Elliptic Curves such as the well-known curve 25519 [Bernstein06].

We assume the existence of a cryptographic hash function $M$ which, on input from $\{0,1\}^*$, outputs a random element from the group $G$. Constructions for such hash functions towards groups such as $G$ exist, such as e.g. Elligator [BHKL13].

*Secret Sharing*

We use a concept called Shamir's Secret Sharing [Shamir79]. It is parameterized by a number of parties $n$, a modulus $p$ and a threshold $0 < t \leq n$. Further, Shamir's Secret Sharing uses an algorithm  which, on input $(n, p, t)$ as well as a secret $x \in Z_p$ creates shares $x_1, \ldots, x_n$ such that:

1. Anyone possessing fewer than $t$ shares from $\{x_1, \ldots, x_n\}$ has no information about $x$.
2. Anyone possessing $t$ or more shares of $\{x_1, \ldots, x_n\}$ can reconstruct $x$.

In Shamir's Sharing, the  algorithm can be instantiated using polynomial arithmetic as follows:

---

[1] We consider the curator to be only an abstract entity and not a concrete organization. This will become more clear in the description below.

1. To share the secret , $x \in Z_p$ sample $t-1$ values $x_1, \ldots, x_{t-1}$ uniformly from $Z_p$.
2. Compute the unique monic degree-$t-1$ polynomial $f(X)$ with coefficients from $Z_p$ where $f(0) = x$ and $f(i) = x_i$ for $i = 1, \ldots, t-1$.
3. Define $x_j = f(j)$ for $j = t, \ldots, n$

*Lagrange Interpolation*

Given any $t$ shares (for simplicity, $x_1, \cdots, x_t)$ there can only exist one monic polynomial $f$ of degree $t-1$ with coefficients over $Z_p$ such that $f(i) = x_i$ for $i = 1, \ldots, t$ by the fundamental theorem of Algebra. Using so-called Lagrange interpolation one can, using a set $L$ of $t$ evaluation points (in our example, $L = \{1, \ldots, t\}$) as well as an additional index $h$, compute coefficients $\lambda_1^{L,h}, \ldots, \lambda_t^{L,h}$ which are also from $Z_p$. Then, for these coefficients it must hold that $f(h) = \sum_{i \in L} \lambda_i^{L,h} x_i$ over $Z_p$. This means, one can evaluate the unique polynomial $f(X)$ in any point by computing a linear combination of any $t$ evaluation points of $f$, and where the coefficients of the linear combination only depend on $L$, $h$ but are independent of the concrete polynomial $f(X)$.

*What is a DOPRF (in our setting)?*

A DOPRF is an interactive protocol run between a client and $n$ key holders. The client has an input $x \in \{0,1\}^*$ while each key holder has as input a Shamir share $k_i$ of the key $k$. At the end of the interactive protocol, the client has learned a value $y \in G$ while no keyserver learns anything about $x$ or $y$. Furthermore, the client learns nothing about $k$. Finally, for the value $y$ it holds that it is uniformly random in $G$ to anyone who does not know $k$, meaning that it reveals no information about $x$.

*How is the DOPRF used in SecureDNA?*

Upon system initialization, a random key $k$ is chosen centrally and then secret-shared using the Shamir's scheme:, each of the $n$ keyservers obtains a share of $k$, with keyserver $i$ obtaining the share $k_i$. After this initial phase, the key $k$ is securely deleted from the place where it was generated, such that only remaining information about it are the shares held by the keyservers.

After this initial phase is completed, the system is operational. The database server starts with an initially empty table $H$. To add values to it, the curator uses the DOPRF with inputs $d_1, \ldots, d_J$ together with the keyservers, obtaining the DOPRF outputs $hd_1, \ldots, hd_J$. It then sends $hd_1, \ldots, hd_J$ to the database server, which adds $hd_1, \ldots, hd_J$ to $H$ This process can be repeated as often as necessary.

To test if a sequence is dangerous, the synthesizer will first compute $s_1, \cdots, s_I$ from the sequence and apply the DOPRF together with the keyservers to any $s_i$, computing hashes $hs_1, \ldots, hs_I$. It then sends the $hs_1, \ldots, hs_I$ to the database server, which informs the synthesizer if any $hs_i$ shows up in $H$.

*On the Security of SecureDNA*

We assume that parties in SecureDNA only try to learn secrets from normally running the involved protocols, but they never actively deviate from the algorithms as specified[2]. This is the ``semi-honest" security model, which is well-established in cryptography. Given that the DOPRF has such semi-honest security itself, the same holds for our overall construction: the curator will not learn any information about $k$ due to the security of the DOPRF, and the same holds for the synthesizer. At the same time, the keyservers learn no information about what the database is curated with or what a customer orders based on the security of the DOPRF. Finally, the database server never even talks to the keyservers, and the only values that it sees are uniformly random outputs of the DOPRF which reveal nothing about the inputs by assumption.

---

[2] We can strengthen our security model to tolerate active corruption of the keyserver and database server. For the keyserver, one can protect each DOPRF computation using standard Zero-Knowledge proofs of exponentiation. To tolerate active corruption of the database server, one can replicate the database across multiple servers which are contacted by the synthesizer, who then uses the majority vote on their responses. For example, 3 copies can tolerate one corrupted server.

Additionally, we also have strong security of $k$: any $t-1$ or fewer corrupted (e.g. hacked) keyservers do not have enough information to recover $k$, based on the guarantees of the Shamir Sharing.

*How the DOPRF is realized*
SecureDNA uses a version of the NPR DOPRF (see [NPR99]):

1. On input $x \in \{0,1\}^*$ the client first chooses a uniformly random $\beta \in Z_p$ and computes $X = M(x)^\beta$. It then chooses a set $L \subset \{1, ..., n\}$ of $t$ keyservers and sends $(X, L)$ to each keyserver in $L$.
2. Each keyserver in $L$, upon obtaining $(X, L)$ from the client, computes $\lambda_i^{L,0}$ as well as $Y_i = X^{k_i \cdot \lambda_i^{L,0}}$ and returns it to the client.
3. Upon obtaining all $t$ responses from the keyservers in $L$ the client computes and outputs $Y = (\prod_{i \in L} Y_i)^{\beta^{-1}}$

Here, the multiplication with $\beta$ and $\beta^{-1}$in the exponent cancels out while the Lagrange coefficients interpolate the shares in the exponent, leading to the output being $Y = M(x)^k$. The value $Y$ can be shown to be indistinguishable from a uniformly random element in the group $G$, due to the Decisional Diffie Hellman assumption assumed to hold in $G$. We refer to [NPR99] for more details about the security.

Observe that this version of the protocol is optimized for low computation on the client-side, as it outsources the application of the Lagrange coefficients to the servers while requiring the client to always obtain responses from all $t$ keyservers in $L$. In case a keyserver $i$ does not respond (e.g., downtime), the protocol can simply be restarted with a new set $L$ that does not contain $i$.

*Additional protection mechanisms for the key*
In the final SecureDNA construction in production, we actually do not generate the key $k$ centrally but use a distributed protocol such that key shares of $k$ can be generated while $k$ never appears on any machine. We additionally use this protocol to generate new keys $\tilde{k}$, and cryptographic protocols that allow us to generate an update key to update $H$ from $k$ to $\tilde{k}$ as well as a protocol for performing this update (without ever revealing the update key or $H$ in the process). Moreover, we use so-called proactive secret sharing mechanisms which regularly redistribute $k$ among the keyservers by generating new shares through a special cryptographic protocol. This means that, if someone steals key shares at time $i$, those cannot be used to reconstruct $k$ any more after the key redistribution protocol was run, nor can any shares which span a redistribution event be used to reconstruct any keys. All of these protocols follow well-established design patterns for multiparty cryptography, such as outlined in [CDN15].

References for Appendix C

## Appendix D: Deployment Cost Analysis

This appendix provides an assessment of the costs associated with implementing the SecureDNA system under various deployment scenarios. We analyze three primary deployment approaches: cloud-based virtual servers, local physical hardware, and colocation facility deployment.

Deployment Options and Cost Considerations.

*Cloud-Based Virtual Servers*

Cloud-based deployment allows providers to utilize virtual server instances only when needed, incurring costs proportionate to usage time. Major cloud providers typically charge per hour of operation, with costs scaling linearly with computational resources. For example, a 32-core/64-thread virtual server costs approximately $2.464 per hour on-demand, amounting to $15,700 annually if operated continuously. Some providers offer annual reservations at approximately half the on-demand cost, but these are only economical for continuous, year-round usage.

*Local Physical Hardware*

Local hardware requires an initial capital expenditure followed by operational costs. A mid-range server (16-core/32-thread) has an approximate acquisition cost of $2,000, with additional power consumption costs of $200-300 annually depending on local electricity rates. Over a standard five-year depreciation period, the total cost amounts to approximately $3,000, significantly less than comparable cloud solutions. Higher-performance servers (e.g., systems with 32+ cores) increase capital costs to $5,000-10,000, with proportionally similar operational expenses.

*Colocation Facility Deployment*

Colocation places provider-owned hardware in professionally managed data centers. This approach maintains the same capital expenditure as local hardware but adds facility fees of approximately $50 monthly per server ($600 annually). While this increases the five-year cost to approximately $5,000 for a mid-range server, it provides enterprise-grade power reliability, cooling, and network connectivity that may not be feasible in local deployments.

*Comparative Cost Analysis*

When comparing deployment options across a five-year period, the economics heavily favor physical hardware for consistent workloads. A mid-range local server costing $3,000 over five years delivers equivalent computing power to a cloud-based server that would cost approximately $39,000 over the same period—a cost difference of 13-fold. However, cloud solutions offer advantages for variable workloads, where servers can be deactivated during periods of inactivity.

Performance Benchmarks.

The SecureDNA synthesis client ("synthclient") achieves screening rates of 1,200-1,300 base pairs per second per CPU thread on typical server hardware. This translates to approximately:

- 49,500 base pairs per second on a mid-range 16-core/32-thread server
- 81,600 base pairs per second on a high-end 32-core/64-thread server
- 572 base pairs per second on a low-cost embedded device ($50 Raspberry Pi 4)

Deployment Recommendations.

*For Commercial Providers*

High-volume providers (processing 40 million base pairs in under 10 minutes, several times daily) have two economical options:

- Cloud-based approach: Using a 32-core virtual server costs approximately $0.41 per 40-million-base-pair screening, resulting in a cost of approximately $1,300 annually for ten such screenings daily.
- Dedicated hardware approach: A high-performance physical server costs $5,000-10,000 in capital expenditure but becomes more economical for frequent screening operations over a multi-year period.

Standard providers (processing primarily gene-length orders) can achieve cost-effective operation even with modest hardware. A $2,000 mid-range server can screen at 43,000 base pairs per second, completing typical gene orders in seconds, while costing significantly less than a continuously operating cloud server over a one-year period.

*For Benchtop Synthesizers*

For on-site benchtop synthesizers, even basic computing hardware such as a Raspberry Pi can process typical synthesis runs in under one minute, making it a highly cost-effective approach for local screening.

*For the SecureDNA System Infrastructure*

The central SecureDNA system infrastructure is most economically deployed using physical servers in colocation facilities. This approach balances cost-effectiveness with the high reliability and performance requirements of the system.

For handling variable load profiles, a hybrid approach may be optimal:

- Small, continuously operating cloud servers or physical hardware for baseline screening demand
- Larger, on-demand cloud servers that activate automatically to handle periodic high-volume screening requests

This analysis demonstrates that while cloud-based deployment offers flexibility, the economics significantly favor physical hardware for steady-state operations, with potential cost savings of more than 90% over multi-year periods.

## Appendix E: Measures for Customer and Provider Data Privacy

This study utilizes unfiltered customer sequence data from specified time intervals provided by multiple DNA synthesis companies. Acknowledging legitimate privacy concerns, this Appendix aims to assure readers as well as the data providers and their customers that stringent measures were taken to fully protect sensitive information in the analysis. The identities of customers were anonymized and not known to the researchers, preventing any possibility of directly revealing them accidentally or intentionally. Only the rates of detected controlled sequences and approximate dataset sizes are reported, with no other specific sequence information disclosed, such that the only sequence content revealed pertains to regulated pathogens. We utilize data from three or more customers per provider and three or more providers total. If there were only two customers per provider, one customer could attribute controlled sequences exclusively to the other. However, with data from at least three customers included, no single customer can definitively attribute a detection to a specific other customer. Similarly, if there were only two customers, one could roughly estimate the order volume of the other by subtracting their own volume from the approximate total given. But with three or more customers aggregated, no single customer can determine the exact order volumes of other specific customers, only an upper bound on their own fraction of their provider's business, a quantity that could be derived from public sources like earnings reports. Regarding the privacy of contributing providers, we can reasonably assume that each contributing provider knows the contents of the dataset contributed to the analysis, allowing them to potentially deduce which data originated from other providers. However, with data incorporated from at least three different providers, no single provider can definitively attribute a particular dataset to another specific provider. To further obscure which controlled sequences may be attributable to which provider, and to confirm that screening is working as intended even in low-hazard-content sets, we intentionally inject known controlled sequences into test data. Since only lower bounds on dataset sizes are provided, reporting controlled sequence detection rates with low precision ensures that any small rates are indistinguishable from zero. Thus, a reported rate of <0.1% could reflect an arbitrary number of actual detections in the provider data, including zero, preventing positive identification of which provider's data may have contained specific controlled sequences. In summary, by utilizing data from three or more customers per provider and at least three providers, along with intentionally limiting the precision of the reported controlled sequence rates and dataset quantifications, this methodology aims to effectively evaluate the screening system while stringently protecting the privacy of all data providers and their customers.